\documentclass[prb,preprint]{revtex4-1}
\usepackage{graphicx} 
\usepackage{float}
\usepackage{appendix}
\usepackage{amsmath}
\DeclareMathOperator\atanh{atanh}
\usepackage{subcaption}
\usepackage{amsfonts}
\usepackage{xcolor}
\usepackage{comment}
\usepackage{bm}
\usepackage{ulem}

\newcommand{\argmax}{\mbox{argmax}}
\usepackage{hyperref}
\usepackage{placeins}

\begin{document}

\title{Learning and Testing Inverse Statistical Problems \\ For Interacting Systems Undergoing  Phase Transition}
\author{Stefano Bae$^{1,2}$, Dario Bocchi$^{1,2}$, Luca Maria Del Bono$^{1,2}$, Luca Leuzzi$^{2,1,*}$}
\affiliation{$^1$ Physics Department, Sapienza University of Rome, Piazzale Aldo Moro 5, 00185, Rome, Italy}
\affiliation{$^2$ Institute of Nanotechnology, National Research Council of Italy, CNR-NANOTEC, Rome Unit c/o Sapienza University of Rome, Piazzale Aldo Moro 5, 00185, Rome, Italy}
\email{luca.leuzzi@cnr.it}
\date{\today}

\begin{abstract}

Inverse problems arise in situations where data are available, but the underlying model is not. It can therefore be necessary to infer the parameters of the latter starting from the former. But, what happens when the model undergoes a phase transition? That is, if, depending on the parameters, data are of different nature? Statistical mechanics offers a toolbox of techniques to address this challenge. In this work, we illustrate three of the main methods: the Maximum Likelihood, Maximum Pseudo-Likelihood, and Mean-Field approaches. We begin with a pedagogical  theoretical introduction to these methods, followed by their application to inference in several well-known statistical physics systems undergoing  phase transitions. 
We consider the bond-ordered and disordered Ising models, the vector Potts model, and the Blume-Capel model on both regular lattices and random graphs. This discussion is accompanied by a GitHub repository that allows users to  produce synthetic data, reproduce the inference results and easily experiment with new systems.

\end{abstract}

\maketitle

\section{Introduction}


 Inverse problems in statistical physics are widely studied to deal with inference and learning on big (but also moderate-dimensional) data systems. From experimental measurements of the system variables,   or from their averages and correlations, it is possible to quantitatively infer the values of the relevant parameters of the statistical physics theory underlying the behavior of the system.
    The term {\it inverse} denotes that the estimation of model parameters is the inverse procedure with respect to the standard applications of statistical mechanics.
    The standard (\textit{direct}) problem, indeed, is to use the theoretical model to predict the behavior of thermodynamic observables, like, e.g., energy, magnetization or specific heat, and, in general, any measurable functions of the dynamic variables. 
    The values of the {\it external} parameters involved in the model definition, like, for instance, magnetic exchange couplings, magnetic fields, crystal fields, chemical potential, temperature, etc.
   are given and the collective properties of the system variables are derived. On the contrary, in the inverse problem one aims at inferring these quantities starting from observations. One can easily identify the practical utility: in many real-world situations, one can obtain fairly easily a set of data by simply performing multiple measurements. The challenge, then, becomes extracting meaningful information from them, for instance in the form of a set of parameters defining a model which one assumes is at the base of the data generation.


   In the era of artificial intelligence one might want to provide a brief explanation on what we mean by statistical inference and how this might differ from a machine learning approach. 
Statistical inference and machine learning are deeply intertwined fields that share common roots in probability and data analysis, yet they often emphasize different objectives. At their core, both aim to extract meaningful insights from data, but they do so with somewhat distinct priorities.
Statistical inference traditionally focuses on understanding the underlying data-generating process, rigorously testing hypotheses, and quantifying uncertainty (the {\it errors}).
It focuses on interpretable models where parameters have clear meanings, and assumptions about the data (like the distribution properties, e.g., the equilibrium Boltzmann-Gibbs distribution in the following cases) are carefully validated.
Machine learning, on the other hand, prioritizes predictive accuracy and generalization to unseen data. 
While it borrows many techniques from statistics — such as regression, Bayesian methods, and likelihood estimation — it often employs more flexible, complex models (like multi-layer {\it deep} neural networks) that may sacrifice interpretability for performance. 
Machine learning is particularly effective in settings, such as image recognition or natural language processing, where traditional statistical models might struggle.

Let us take, for instance, the noisy functional relationship between {\it input} variables  $\bm x$ and {\it output} variables $\bm y$:
$$\bm y = F(\bm x| \{\lambda\}) + {\bm{\xi}}, $$
where $\xi$ is a generic stochastic noise.
In statistical inference we propose a theory modeled by a given function $F$ and estimate the parameters $\{\lambda\}$ of $F$ such as those values that best reproduce the measured values of the output dataset $\bm y$ given the measured values of $\bm x$. 
A common drawback occurs when the model $F$ poorly represents the phenomenon and the inference may appear to produce good parameter estimates$\ldots$ but for the wrong theory.
In machine learning, we train a generic function $F$ to reproduce the outputs $\bm y$ by feeding it the input dataset $\bm x$ through some algorithm. The usual approach is to use \textit{gradient descent} (GD), or one of its more advanced versions, which is an iterative method that updates the function parameters according to
\begin{equation}
\lambda_{t+1} = \lambda_t + \eta \ \nabla_{\lambda} \mathcal{L}\left(F(\bm{x} \mid \{\lambda_t\}), \bm{y}\right),
\label{GD}
\end{equation}
where $\lambda_t$ denotes the parameters at iteration $t$, $\eta$ is the \textit{learning rate} (LR), and $\mathcal{L}$ is a suitable \textit{loss function}. It is a function that quantifies the discrepancy between the model predictions and the target data. Obviously, we want to minimize it. Each iteration $t$ is called an \textit{epoch} and corresponds to a complete pass over the training dataset. The learning rate $\eta$ controls the step size in parameter updates: a value too large may cause divergence and make the procedure unstable, while one too small may make convergence slower. To improve stability and convergence, one technique is to progressively decrease the learning rate during training. For instance, an exponential decay of the learning rate can be carried out via a \textit{decay factor}, such that $\eta_t = \eta_0 \cdot \text{(decay \,factor)}^{t}$. 

A key limitation of the machine learning approach is \textit{overfitting}: different functions $F$ may satisfy the same equation $\bm y = F(\bm x)$ during training — but which one is correct on arbitrary (unseen) data? In machine learning  one wants the learned function to work well with out-of-sample, previously unseen data. That is, in learning jargon, a good   {\it generalization} is required. A function with too many parameters (i.e., {\it over}fitting) might precisely adapt to the training dataset but generalize very badly. This is the well known overfitting vs. generalization dilemma. 

Despite these differences, the boundaries between the statistical inference and machine learning are not strict and even increasingly blurred as extensive research in the framework of artificial intelligence exponentially develops. 
In essence, statistical inference provides the theoretical foundation for reasoning about data, while machine learning extends these ideas to build systems that learn from data at large scale. The interplay between them continues to evolve, with each field enriching the other.

\subsection{Structure of the paper}
  
   In this pedagogical paper we focus on  various inverse problems whose direct versions undergo {\it phase transitions}. That is, the collective behavior of the system qualitatively changes, more or less abruptly, 
at some critical values of the external parameters.
Before dealing with the  inference, in Sec. \ref{Sec:II}, we, therefore,  briefly report the salient features of some {\it direct problems}.  In order to be able to illustrate how   inference works under different circumstances, we 
 span  across models displaying different kinds of phase transitions. 
As explanatory instances we chose to work with 
\begin{itemize}
    \item 
the \textbf{Ising model} \cite{ising1924beitrag}, i.e., a model of axial (binary) spins, which undergoes a Curie-Weiss ferromagnetic transition, and its bond-disordered version, yielding a spin-glass phase  at low temperature;
\item 
the \textbf{Potts clock model} \cite{RevModPhys.54.235} for discrete planar spins, which undergoes either a first order or a second order ferromagnetic transition, depending on the number of discrete spin states;
\item the \textbf{Blume-Capel model} \cite{B66,C66} for magnetic systems with single-ion anisotropy in which a tri-critical point occurs between first and second order phase transitions, having parallels in the$\phantom{!}^4$He superfluid transition.
\end{itemize}

\subsection{An introductory track }
For the reader that is less familiar with the topic, or statistical mechanics in general, we structured the paper so that they can easily select only the parts related to the \textbf{Ising model}, as they provide a standard introduction to the used methods, and skip some technical details and more mathematical derivations. The interested reader can then expand by considering also the \textbf{Potts clock model} and the \textbf{Blume-Capel model}, as well as the more technical aspect about the inference methods and their connections to machine learning issues. We have identified the paragraphs that can be skipped with a star symbol $\mathrm{(}\star\mathrm{)}$: these sections can be safely skipped on the first read, without losing the main points of the topic.

\subsection{Synthetic data production}
In order to  produce synthetic data, we have performed Markov Chain Monte Carlo numerical simulations of the dynamics of the above mentioned systems. In particular, we have used either the Parallel Tempering algorithm, else termed the Exchange Monte Carlo \cite{marinari1992simulated,Hukushima}, or the Wolff algorithm \cite{wolff1989collective}, depending on the model features  \footnote{We refer the interested reader to \cite{newman1999monte} for a pedagogical introduction to these algorithms.}. Data are generated at equilibrium, i.e., ensembles of well thermalized spin configurations uncorrelated in time, and they are used as input for the inference of the system coupling parameters.
We have been considering both models on 2D and 3D nearest neighbour lattices and on Erd\"os–Rényi (ER) random graphs.
The codes for the Monte Carlo simulations for the Ising (2D, ER), the Potts clock (3D, ER) and the Blume-Capel (2D, ER) models are available at the GitHub repository \url{https://github.com/bsfn-0323/inverse_ising}.

\subsection{Summary of contents of inference}

After an introduction on the Likelihood method (Sec. \ref{sec:secIII}), in Sec. \ref{Sec:MAX_PL}-\ref{INVERSE_MF} we  focus on two widely spread inference methodologies for systems with interacting variables, that are relatively easy to implement and understand and are both based on equilibrium statistical mechanics:  the Mean-Field approach (Sec. \ref{INVERSE_MF}) and  the Pseudo-Likelihood method (Sec. \ref{Sec:MAX_PL}). The first one relies on the measurements of averages and two-point correlations and reconstructs the coupling matrix as the inverse of the covariance matrix of the data\cite{Opper2001}. The second one starts from considering a factorized partition function on the spins in which all data but one at a time are fixed at the measured values. This turns out to be equivalent to multinomial logistic regression in statistics, in which the threshold of the sigmoid function is represented by the energy contribution of the free variable \cite{Ravikumar2010}.

In Section \ref{Sec:SI} we report the outcome of both Mean-Field and Pseudo-Likelihood analysis to the various models for which we have produced synthetic data.

 Applications of these techniques, or their more complicated (but not so exceedingly better performing) derivatives and generalizations, dealing with the learning of the model parameters, are many and extremely interesting \cite{Opper2001,nguyen2017inverse}. They span a wide range of disciplines and include  
 reconstruction
of synaptic connections and neuronal activations in biological neural networks \cite{schneidman2006weak, potthast2022inverse, pesaran2018investigating},  determination of protein structure and folding \cite{burger2010disentangling, anishchenko2018contact, cocco2018inverse},  inference of gene regulatory
networks \cite{wang2022inference, friedman2004inferring}, retrieval of transmission properties of light across random media, image reconstruction or focusing,\cite{Popoff2010,lucas2018using,bertero2021introduction,Ancora2022,Ancora2022b}, reconstruction of gain competition in random lasers\cite{Tyagi2016,Marruzzo2017,Marruzzo2018} and of effective interaction in linear wave systems\cite{Tyagi2015}, prediction of epidemic spreading and  detection of patient zero \cite{shah2020finding, biazzo2022bayesian},  medicine \cite{bergquist2023uncertainty}, routing optimization \cite{Xu2023}, minimization of risks in financial investments and stock market analysis \cite{bury2013market, zhao2018stock}, social networks properties \cite{shumovskaia2021online}, among others.

\section{The direct problems}
\label{Sec:II}
We now move to describing in detail what models we consider in this paper and what are the corresponding direct problems.
\subsection{Ising model}
The Ising model, first introduced in 1925 by the physicist Ernst Ising \cite{ising1924beitrag}, is often considered as the cornerstone for the study of phase transitions and critical phenomena in statistical mechanics. This model provides a powerful framework for describing the magnetic behavior of a collection of spins in a material. In the Ising model, each microscopic intrinsic atomic angular momentum, or {\it spin}, is represented by a binary variable $s_{i} = \pm 1$, that can take on two values, typically denoted as {\it up} or {\it down}. The exchange magnetic interactions between spins are captured by an interaction matrix $J_{ij}$ that favors alignment (ferromagnetic interactions) if $J_{ij} > 0$ or anti-alignment (antiferromagnetic interactions) if $J_{ij} < 0$. When couplings are symmetric and the system is kept at a certain temperature $T$, it can reach an equilibrium steady state described by the \textit{Boltzmann-Gibbs} distribution $P_\text{BG}$,
\begin{equation}
\label{GB_DISTR}
    P_\text{BG}(\{s\}) = \frac{e^{-\beta\mathcal{H}(\{s\})}}{Z}
\end{equation}
where $\{s\}$ is the set of all the spins, $\beta = 1/T$ is the inverse temperature and $Z$ is the normalization function,  called the \textit{partition function} 
\begin{equation}
 Z= \sum_{\{s\}}e^{-\beta \mathcal{H}(\{s\})}.
\end{equation}
The model \textit{Hamiltonian} $\mathcal{H}$ is defined as:
\begin{equation}\label{eq:ising_ham}
    \mathcal{H}(\{s\}) = -\sum_{\langle i j \rangle} J_{ij}s_{i} s_{j} \, ,
\end{equation}
where $\langle i j \rangle$ represents the sum over all distinct connected pairs. The choice of how to structure connections in a model of interacting variables can take on various topologies. A rather interesting and common choice is to use lattice models \cite{orrick2001susceptibility}, where the structure of the connections is completely deterministic and it is usually taken to be {\it nearest neighbors}, i.e., each spin interacts only with the spins nearest to it in the lattice. For simple $D$-dimensional hypercubic lattices this means that each spin has connectivity - also termed \textit{coordination number} - $2D$. 

\subsubsection{The ferromagnetic Ising model on the two-dimensional lattice}
\begin{figure}
    \centering
    \includegraphics[width=0.95\linewidth]{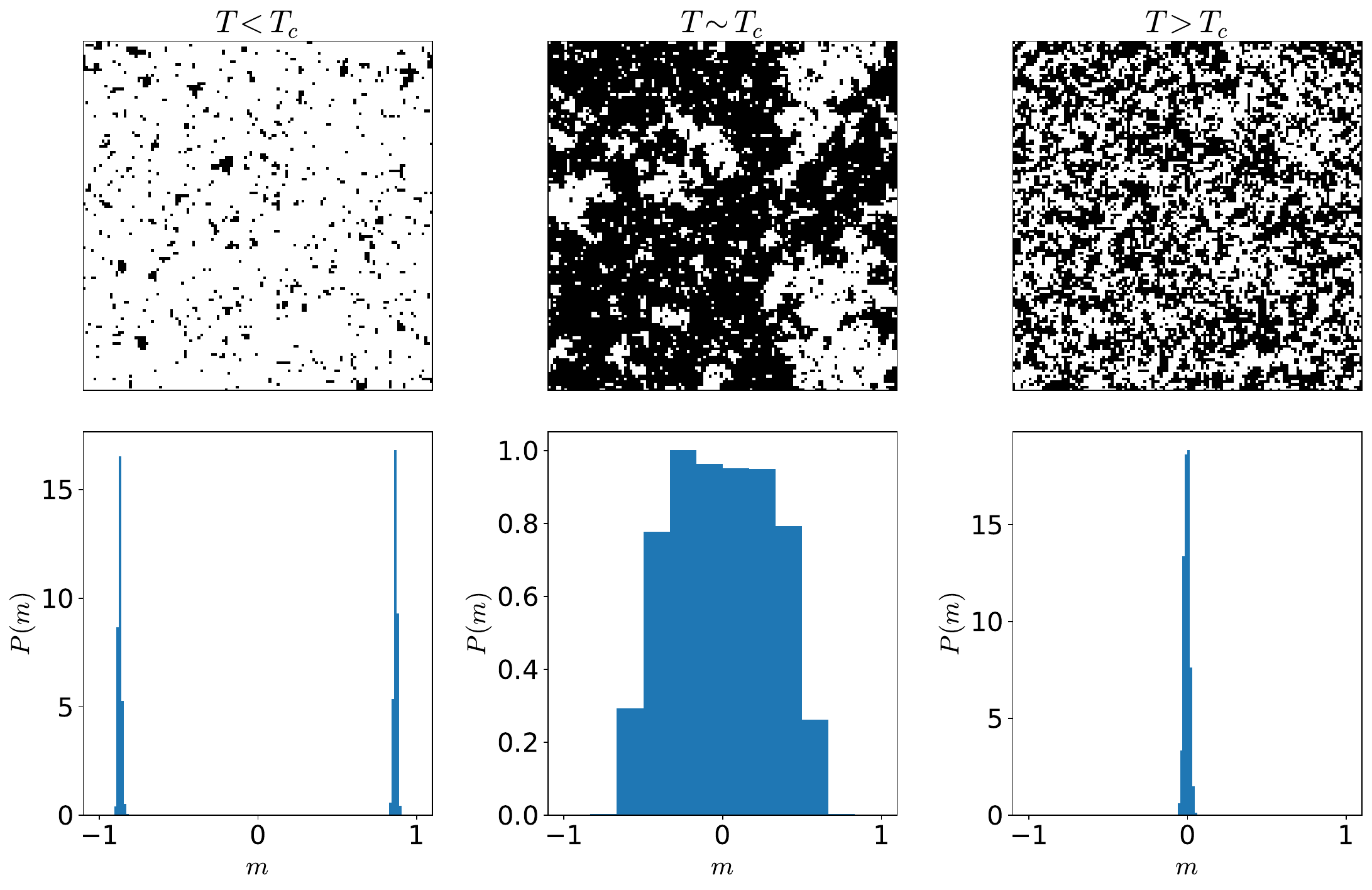}
    \caption{Monte Carlo simulation of the 2D Ising model with $N=128$. The first row displays lattice snapshots (white and black pixels identify $+1$ and -1 spins, respectively)  at $T < T_c$ (left) and $T > T_c$ (right), highlighting the emergence of large-scale clusters near the critical point (center). The second row illustrates the magnetization probability distribution $P(m)$ for $T < T_c$ (left), $T \sim T_c$ (center), and $T > T_c$ (right), showcasing the transition from a bimodal ordered state to a narrow disordered distribution.}
    \label{fig:ising_chapter}
\end{figure}
In order to give more intuition on the physics of the model, let's take for a while the ferromagnetic Ising model defined on a two-dimensional lattice, which is one of the most famous models in statistical mechanics. In this case, all the couplings are taken equal and ferromagnetic, i.e. $J_{ij} = J > 0$.
Depending on the inverse temperature $\beta = 1/T$, the Ising model can be in different physical states. The state is described by \textit{magnetization}:
\begin{equation}
    m = \frac1N\sum_{i=1}^N s_i \,,
\end{equation}
which tells us if the system has some kind of magnetic order. Indeed if $m\simeq0$ it means that the system has no magnetic order, while if $|m|>0$ there is an order caused by the alignment of the spins $s_i$. More generally any kind of macroscopic variable of the system that is used to detect the status of a system is called \textit{order parameter}. But why there is an ordering in the first place? Looking at the Hamiltonian by taking all the $J_{ij}=1$ one can see that the energy will be greater when the product $s_is_j >0$, i.e. more neighboring spins are aligned. In fact, if only the energy is considered the ideal configuration will be to have all spins aligned to $\pm 1$. However, the Boltzmann-Gibbs measure contains also the temperature. For $T\rightarrow 0$, $\beta\rightarrow\infty$ so all the measure will be concentrated close to the minima of the energy function, with a magnetization closely peaked around large values of the magnetization (Fig. \ref{fig:ising_chapter} left column). By increasing the temperature we allow the configurations to have misaligned spins. If the temperature is too high $\beta \rightarrow 0$ the contribution of the energy will be so small that the configuration will be all random giving $m \sim 0$ (Fig. \ref{fig:ising_chapter} right column). At this point the intuition is that there will be a critical value of the temperature $T_c$ for which the magnetization starts to become larger than one, $m\neq0$ (Fig. \ref{fig:ising_chapter} central column). At this value of the temperature the probability distribution of the magnetization becomes very broad. Another way of seeing this is through the connected correlation:
\begin{equation}
    C_{ij} = \langle s_is_j\rangle - \langle s_i\rangle\langle s_j\rangle.
\end{equation}
Let us imagine to fix the spin $s_0$. At infinite temperature the spins behave randomly, therefore on average, the spin $s_0$ behaves differently from any other spins $s_j$ ($s_0s_j$ will have random sign so $C_{0j}\sim 0$). By lowering a bit the temperature, spins $s_j$ close to $s_0$ in a certain radius $r$ will try to align with it while the others still behave randomly. It's intuitive to see that by lowering the temperature $r$ will increase and more spins will behave similarly: we say that there are clusters of radius $r$. At zero temperature, the configurations are frozen, but still $C_{ij} \sim 0$. This is because the connected correlation looks at fluctuations around a certain state: on average $\langle s_i s_j \rangle \sim 1$ but also $\langle s_i \rangle \langle s_j \rangle \sim 1$. By increasing the temperature, we observe fluctuations around this fixed state, and the size of the misaligned clusters will be $r$. In the thermodynamic limit ($N \rightarrow \infty$), this correlation length diverges at a critical temperature, $T_c$. However, in any real or simulated finite system, $r$ cannot exceed the physical dimensions of the lattice. Therefore, as the temperature approaches $T_c$ in a finite system, the correlation length $r$ instead peaks, reaching a maximum bounded by the system size. One can visually see the correlation lengths by looking the first row of Fig. \ref{fig:ising_chapter}: for $T\sim T_c$ the size of connected clusters is large, while for $T>T_c$ are smaller and for $T<T_c$ the size of the clusters different from the mean behavior are also small.\\

\subsubsection{Generalization of the ferromagnetic Ising model on the lattice}

Despite the very interesting phaenomenologuìy that it exhibit, the ferromagnetic Ising model on the lattice is only one of the possible examples of an Hamiltonian in the form described by \eqref{eq:ising_ham}. Indeed, one can take the couplings $J_{ij}$ to be all negative instead of all positive (thus describing an \textit{antiferromagnetic} system) or even randomly distributed.

Additionally, one can consider a generic form of interaction network, instead of considering a $D$-dimensional lattice. For example, other fundamental interaction networks are random graphs, in which a spin has a finite number of connections, but they are chosen at random between all the possible pairs. For these graphs, the notion of distance does not change the critical behavior with respect to networks with all-to-all interactions, and the so-called mean-field theory is not an approximation but holds exactly. In particular, in this work we choose to work with Erd\"os–Rényi  random graphs  \cite{kabluchko2019fluctuations}, that  are very easy to implement. In ER graphs, the connectivity between spins is governed by a Poisson probability distribution. 

The structure of the connections is encoded in a adjacency matrix. Furthermore, each non-zero link can take given values characterizing the model properties. 
In the ordered case, the non-zero $J_{ij}$ elements of the interaction matrix can be taken with uniform values, yielding a constant interaction strength across all spin pairs and translational invariance in ordered lattices. 
These are models used to study, for instance, ferromagnetism. Alternatively, in the bond-disordered case, the value of each element  $J_{ij}$  can be extracted from a random distribution for every spin pair. 
If $P(J_{ij})$ has a  large enough variance,  with respect to its average, at low temperature the system might undergo a transition to a spin-glass phase \cite{parisi2023nobel}. On the contrary, when couplings are random but mostly positive, one has a random ferromagnet in the cold phase.

We implemented the study of four distinct cases for the systems with Ising variables to provide synthetic data to be later used to feed the inference procedures:
\begin{itemize}
    \item  The square lattice with ordered couplings ($J_{ij} = J = 1$ for all nearest neighbors);
    \item the square lattice with very disordered  couplings, i.e.,  distributed with a normal distribution $\mathcal{N}(0, 1)$ of zero mean and unit variance;
    \item the Erdős–Rényi random graph $G_N(p)$ with connectivity $c$. This is a graph in which  each one of the $N(N-1)/2$ possible links between $N$ nodes is present with probability $p$, and for which, therefore, the probability distribution of the connectivity comes out to be Poissonian with average connectivity $c=pN$. Coupling values are taken all equal to $1$, we choose $c=4$, i. e. $p=4/N$;
    \item the Erdős–Rényi random graph $G_N(p)$ with connectivity $c=4$ with random, normally distributed, coupling values. 
\end{itemize}  
Three out of four of the models above display a phase transition at a finite temperature, called {\it critical} temperature, below which they acquire a collective behavior that is a ferromagnet in the ordered cases and a quenched disordered spin-glass in the disordered ER graph case. 
The 2D Ising model with random couplings (of zero average), also called the 2D Edwards-Anderson model \cite{edwards1975theory}, does not display a $T_c>0$ phase transition. The system stays in the paramagnetic phase at all $T$ \cite{bray1987scaling}.

The critical temperature of models on ER graphs  can be computed exactly, in the framework of mean-field theory. For the ferromagnetic ER lattice of average connectivity $c$ it is obtained solving the equation  
\begin{equation}
c\, \tanh(J/T_c)=1,
\label{Tc_ER_FM}
\end{equation}
where $J=1$ is the constant coupling between nearest neighbors.
For $c=4$,  it yields $T_c=3.912$. This value can also be obtained using the Belief Propagation approach described in App.~\ref{sec:BP}.
For the spin-glass ER lattice of average connectivity $c$ it is given by the equation 
\begin{equation}
    c\, \mathbb E_J\left[\tanh^2(\sigma_J/T_c)\right]=1,
    \label{Tc_ER_SG}
\end{equation}
where now the expectation value over the $J$ coupling distribution, $\mathbb E_J[\ldots]$ has to be taken. In the  case of  normal distribution 
$P(J)=\mathcal{N}(0, \sigma_J=1)$ and for $c=4$ 
the critical temperature comes out to be  $T_c \simeq1.524$. 
At variance with the random graph cases, the computation of the critical temperature for the ferromagnetic model on the 2D lattice cannot be carried out with mean field techniques and requires a more complicated approach. The first computation was performed by Lars Onsager in his famous paper \cite{onsager1944crystal}, yielding a value $T_c \simeq 2.269$.

In Table \ref{tab:ising_tc}, the critical temperatures for these four case  models are reported \cite{onsager1944crystal, baxter2016exactly,leone2002ferromagnetic, xu2017dynamic, viana1985phase}.

\begin{table}[h!]
\centering
\begin{tabular}{|c|c|c|}
\hline
$T_c$  & Ordered couplings & Disordered couplings \\ \hline
\hline
Square lattice & $2.269$ & $ 0$ \\ 
\hline
ER graph ($c=4$)& $3.912$ & $1.524$\\
\hline
\end{tabular}
\caption{Critical temperatures of the Ising model with various topologies and coupling values. The  ER graph has a Poisson-distributed connectivity of average $4$. The spin-glass network has Gaussian-distributed coupling values of unit mean square deviation, $\sigma_J=1$.}
\label{tab:ising_tc}
\end{table}

\subsection{$\mathrm{(}\star\mathrm{)}$ Potts model }
\label{POTTS}
In the \textit{vectorial Potts model}, also known as the \textit{clock model}  \cite{RevModPhys.54.235}, each spin $\vec{s}_i$ is a 2-dimensional vector of unitary norm, oriented in the plane along one of $q$ possible discrete angles: 
\begin{equation}
\label{THETA_POTTS}
\theta_n = \frac{2 \pi}{q}n \quad, \quad n = 0, 1, \ldots, q-1.
\end{equation} The system is described by the Hamiltonian
\begin{equation}\label{eq:hamiltoniana}
    \mathcal{H}(\{\vec s\}) = -J\sum_{\langle i j \rangle}
\vec{s}_i \cdot \vec{s}_j =-J\sum_{\langle i j \rangle}
\cos(\theta_i - \theta_j),
\end{equation} 
where $\langle i j \rangle$ represents, once again,   all connected pairs and $J$ is a system parameter that defines the strength (and the kind) of interaction, exactly as in the Ising model. 
As before, since changing the absolute value of $J$ only corresponds to a rescaling of the temperature, in the  ferromagnetic case $J$ can be taken to be unity without loss of generality.

When $q = 2$ the spins become Ising variables and the Potts and Ising models coincide. In the $q \to \infty$ limit, spins become continuous on the circumference and  yield the so-called \textit{XY} model which, in 2D, has been of fundamental importance in studying topological transition \cite{kosterlitz1973ordering, kosterlitz2017nobel}. In this work we will consider the $q = 4$ colors case and ordered ferromagnetic transitions.

We consider two distinct networks: a cubic lattice in three dimensions and a $G_N(M)$ Erdős–Rényi graph, which is the ensemble of graphs of $N$ nodes linked by $M$ edges. The slight difference with the $G_N(p)$ case, is that in a $G_N(M)$ Erdős–Rényi graph the total number of links is fixed to be $M$. Notice that the two models are equivalent in the thermodynamic limit, as long as $M = p N(N-1)/2 \simeq c N/2$, i.e, the average number $c$ of links per spin in the $G_N(p)$ case is equal to $c=2 M/N$. Here $M$ is chosen in such a way that $c = 6$,  i.e., that the random graph has the same average connectivity as the cubic lattice.

For the cubic lattice, the critical temperature is numerically determined. Indeed, as in the case of the Ising model, a 3D exact solution does not exists and the mean-field approximation is only  valid at higher dimension. The result of Ref. \cite{scholten1993critical} is $T_c \simeq 2.26$, that is, half the critical temperature of the three-dimensional Ising model \cite{ferrenberg1991critical}. For the ER graph, we find a  critical temperature of approximately $T_c \simeq 2.97$, as shown in App. \ref{app:Tc_potts} using the Binder cumulant approach and in App. \ref{sec:BP} using the Belief Propagation approach.

\subsection{$\mathrm{(}\star\mathrm{)}$ Blume-Capel model }
\label{BLUME}
The third instance of a statistical mechanics model that we are going to focus on  is the Blume-Capel model. It can be considered another generalization of the Ising model, yielding new collective properties, such as a first order phase transition in a given region of the phase diagram. It was first proposed by M.Blume\cite{B66} and H.W. Capel\cite{C66}, to model the magnetic first-order phase transition in uranium dioxide, $\text{UO}_2$. The Hamiltonian of the system is:
\begin{equation}
    \mathcal{H}(\{s\}) = -\sum_{\langle i,j \rangle} J_{ij} s_i s_j + \mu \sum_{i} s_i^2
    \label{eqn:bc_ham}
\end{equation}
\begin{figure}[t!]
    \centering
    \includegraphics[scale = 0.65]{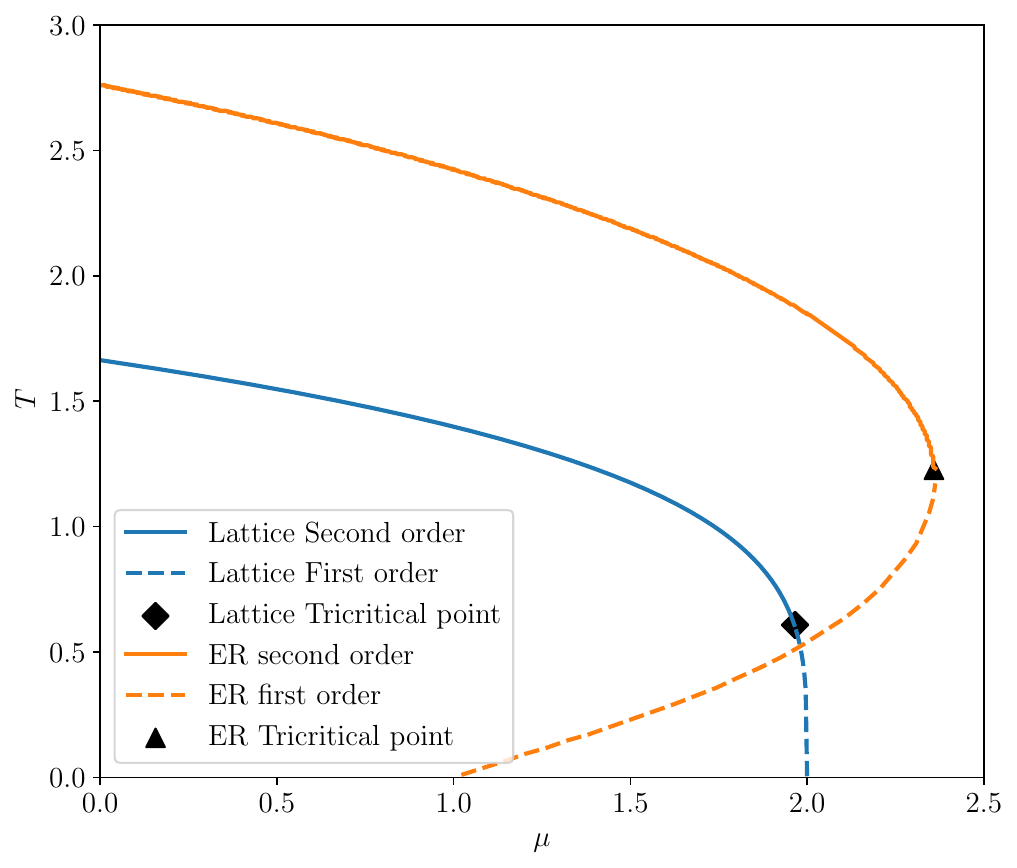}
    \caption{$\mathrm{(}\star\mathrm{)}$ Phase diagrams of the  Blume-Capel models on short range 2D lattice (orange continous and dashed lines) and on the Erdos-Renyi random graph with average connectivity $4$ (blue continous and dashed lines). 
    For the 2D model the orange continous line marks a second order transition between a ferromagnetic phase at low temperature and a paramagnetic one at high temperature. At the tricritical point marked by a diamond the transition becomes first order (orange dashed line). The lines are obtained by finite size scaling of equilibrium data generated by Exchange Monte Carlo algorithm, see Apps. \ref{app:1},\ref{app:2}.
    For the ER model, the second order transition phase transition is displayed as a blue continous curve and the first order transition as a blue dashed curve. 
     These are analytically 
     obtained 
     using 
     Belief Propagation, 
      see App. \ref{sec:BP}. The reentrance at low temperature is typical of many Blume Capel models, it is called inverse freezing or melting \cite{schupper04,crisanti05,schupper05,leuzzi11} and it is responsible for the dynamical effect known as {\it shear thickening}.\cite{sellitto05}}
    \label{fig:bc_phasediag}
\end{figure}
where the spin variables can now take values $s_i =  \{-1,0,+1\}$ and $\mu$, called {\it crystal field}, acts as a chemical potential for the {\it empty species}, i.e., the occurrence of spin $s_i=0$ behaving like non-interacting holes in the graph. Indeed, we can see that a larger $\mu$ gives a larger positive contribute to the energy when $s_i = \pm 1$. 
The sum $\sum_{\langle i,j \rangle}$ is, again, the sum over all couples $i,j$.  When they are nearest neighbors in the simple hypercubic lattice, for example,  interacting $s_i$ and $s_j$ are at distance $1$. 
The model instances that we present here to produce synthetic data for the inference are a 2D square lattice of linear size $L$ with periodic boundary conditions, and a $G_N(M)$ Erdős–Rényi graph.  In particular, we take $M = 2c/N = 2N$ to ensure that the mean connectivity is $c=4$, as in the square lattice case.

In Figure \ref{fig:bc_phasediag} we display the phase diagram of the model when the topology is a 2D lattice or an  Erdos-Renyi random graph. The methods used to characterize the critical points are reported in App. \ref{app:1}, \ref{app:2} and \ref{sec:BP}.

Now that we have introduced the models that we will use to generate data to test the different inference procedures, we now finally move to the inverse problem.

\section{The inverse problem methodology}

\subsection{Max-Likelihood}\label{sec:secIII}

Contrary to the standard framework used in statistical mechanics, that is, to fix a model and then evaluate its observables as functions of the parameters,  inverse problems deal with the inference of the latter using the knowledge of the former. This approach is common in the context of data analysis, where the primary goal is to understand the (unknown) model underlying the (known) experimental data


Let us start from probability distribution of  configurations of spins interacting between them with coupling constants $\{J\}$ and subject to external fields $\{h\}$ according to a Hamiltonian $\mathcal H(\{s\},\{J,h\})$. 
The distribution function in the canonical ensemble 
of an equilibrium statistical mechanics problem at fixed number of spins $N$ and temperature $T=1/\beta$ is the 
previously mentioned Boltzmann-Gibbs distribution (in which  we  explicitly write  the dependence on the set of parameters $J$ ad $h$)
\begin{equation}
    \label{DEF:BG_DIST}
P(\{s\}|\{J,h\}) = \frac{\exp\left\{-\beta \mathcal{H}(\{s\},\{J,h\})
\right\}}{Z(\{J, h\})},
\end{equation}
where the normalization
\begin{equation}
\label{PARTITION}
Z(\{J, h\}) = \sum_{\{s\}} \exp \{- \beta \mathcal{H}(\{s\}| \{J, h\})\}
\end{equation}
is the partition function.
In the direct problem, i.e., computing the properties of a given model, specified by the values of $\{J\}$ and $\{h\}$, one computes the thermal averages at equilibrium of various functions of the spins. For instance the \textit{local magnetization},
\begin{eqnarray}
    \label{MEAN_M}
  m_i =  \frac{\sum_{\{ s\}} \, s_i\, \exp \{- \beta \mathcal{H}(\{s\}| \{J, h\})\}}{Z(\{J, h\})} =  \langle s_i\rangle, 
\end{eqnarray}
the \textit{two-point correlation function},
\begin{eqnarray}
    \label{CORRELATION_C}
  C_{ij} =  \frac{\sum_{\{ s\}} \,s_i\,s_j \,\exp \{- \beta \mathcal{H}(\{s\}| \{J, h\})\}}{Z(\{J, h\})}=
  \langle s_i s_j\rangle 
\end{eqnarray}
and the \textit{two-point connected correlation function} (or \textit{covariance}),
\begin{eqnarray}
    \label{COVARIANCE_G}
  \Gamma_{ij}=  \frac{\sum_{\{ s\}} \,(s_i-m_i)\,(s_j-m_j) \,\exp \{- \beta \mathcal{H}(\{s\}| \{J, h\})\}}{Z(\{J, h\})}=
  \langle s_i s_j\rangle -m_i m_j = \langle s_i s_j\rangle_c
\end{eqnarray}
In the following description we will use Ising spins for illustrative purposes, but everything can be easily generalized to Blume-Capel and Potts-clock spins.

\subsubsection{Posterior and Likelihood}
\label{POST_LIKELIHOOD}
In the inverse problem, the strategy to find  couplings and fields starting from the $\{s\}$ configurations is to maximize the {\it likelihood}. Actually, from a Bayesian perspective \cite{BERNARDO94,DIACONIS18,Leuzzi2025} the probability to be maximized with respect to the $\{J\}$ and $\{h\}$ should actually be the {\it posterior probability}.
Indeed, if the likelihood is $P(\{s\}|\{J,h\})$ and the prior distribution for couplings and fields is $A(\{J,h\})$, then the posterior is the probability distribution of the couplings and fields considering the available measured spin configurations
$\{\bm s\}= \{ \{s^{(1)}\}, \{s^{(2)}\}, \ldots, \{s^{(\mu)}\}, \ldots, \{s^{(M)}\}\}$, with $\mu=1,\ldots,M$ labeling the specific measure out of a total of $M$ measurements. For a single configuration $\{s\}$ Bayes' formula reads 
$$P(\{J,h\}|\{s\}) \propto P(\{s\}|\{J,h\})A(\{J,h\}).$$
For $M$ independent measurements we then get:
\begin{equation}
P(\{J,h\}|\{\bm s\}) \propto \prod_{\mu=1}^{M}P(\{s^{(\mu)}\}|\{J,h\})A(\{J,h\})  .
\label{BAYES_INFERENCE}
\end{equation}
In the case of a likelihood 
$P(\{\bm s\}|\{J,h\}) =  \prod_{\mu=1}^{M}P(\{s^{(\mu)}\}|\{J,h\})$ showing a very pronounced peak in the $\{J,h\}$ space, one usually considers the prior $A$ as constant.
Notice that the prior does not need to be uniform for all the values of couplings and fields, but just in the region of parameter space in which the likelihood is sensibly different from zero. For a large enough number of measurements and processes whose (co)variance is not diverging this is typically the case. Therefore, eventually
$$P(\{J,h\}|\{\bm s\}) = P(\{\bm s\}|\{J,h\}),$$
and maximizing the posterior is equivalent to maximizing the likelihood:
$$\max_{\{J,h\}}P(\{J,h\}|\{\bm s\}) = \max_{\{J,h\}}P(\{\bm s\}|\{J,h\}).$$
Therefore, the above approach is called \textit{max-likelihood}. 

We will not just disregard the prior distribution, though, whose role is fundamental in many less ideal situations\cite{HALDANE48,JEFFREYS48,BERNARDO94,Leuzzi2025}. We will get back to it in Sec. \ref{OVERFITTING}, where it also takes the name of ``regularization''.

In the next (optional) Section $\mathrm{(}\star\mathrm{)}$ we describe for the interested reader how it is implemented in practice. In common cases, however, it is more practical to use approximations, like the max-pseudo-likelihood, which we describe in the successive Section \ref{Sec:MAX_PL}.

\subsubsection{$\mathrm{(}\star\mathrm{)}$ Max log-Likelihood and Boltzmann machine learning }
\label{MAX_LOG_LIKELIHOOD}
Since it is computationally more practical to deal with sums than with products
one can exploit the monotonicity of the $\log$ function and maximize the so-called log-likelihood
\begin{equation*}
\max_{\{J, h\}} \log P(\{\bm s\}|\{J, h\}) = \max_{\{J, h\}} \log \prod_{\mu=1}^M P(\{s^{(\mu)}\}|\{J,h\})
= \max_{\{J, h\}} \sum_{\mu=1}^M \log P(\{s^{(\mu)}\}|\{J, h\})
\end{equation*}
\begin{equation}
= -\max_{\{J, h\}} \sum_{\mu=1}^M \beta \mathcal{H}[\{s^{(\mu)}\}|\{J, h\}] - M \log Z(\{J, h\}),
\end{equation}
rather than the likelihood itself. Here
 we have explicitly written the likelihood in terms of the Boltzmann-Gibbs distribution Eq. (\ref{DEF:BG_DIST}), at the ground of our statistical (or learning) model.  We  now illustrate the detailed computation with the Ising Hamiltonian, Eq. (\ref{eq:ising_ham}), that we rewrite for convenience including the interaction with external local inhomogeneous fields $\{h\}$:
\begin{equation}
\nonumber
\mathcal H(\{s\},\{J,h\}) = -\sum_{i<j}^{1,N}J_{ij} s_is_j-\sum_{i=1}^Nh_is_i. 
\end{equation}
Here the sum runs over all ordered couples of spins, though we imply that $J_{ij}\neq 0$ only for actually interacting couples.
 The max log-Likelihood expression than reduces to 

\begin{equation}
\max_{\{J, h\}} \log P(\{\bm s\}|\{J, h\}) 
= \max_{\{J, h\}} \left[ \sum_{i<j}^{1,N} \beta J_{ij} \sum_{\mu=1}^M s_i^{(\mu)} s_j^{(\mu)} + \sum_{j=1}^{N} \beta h_j \sum_{\mu=1}^M s_j^{(\mu)} \right] - M \log Z(\{J, h\}).
\end{equation}
We can, then, introduce the empirical average of the spin magnetization on each site $j$
\begin{equation}
\label{EXP_M}
\bar m_j = \frac{1}{M} \sum_{\mu=1}^M s_j^{(\mu)}
\end{equation}
and the empirical correlation between spins, averaged over the $M$  measured configurations
\begin{equation}
\label{EXP_C}
\bar C_{ij} = \frac{1}{M} \sum_{\mu=1}^{M} s_i^{(\mu)} s_j^{(\mu)},
\end{equation}
thus leading to the expression

\begin{equation}
\max_{\{J, h\}} \log P(\{\bm s\}|\{J, h\}) = M \max_{\{J, h\}} \left[ \sum_{i<j}^{1,N} \beta J_{ij} \bar C_{ij} + \sum_{j=1}^{N} \beta h_j \bar m_j - \log Z(\{J, h\}) \right]
\end{equation}
Such maximum can be reached by iteration with a machine learning  procedure eventually yielding  the solutions to the equations
$$\frac{\partial \log P(\{\bm s\}|\{J, h\})}{\partial J_{ij}}=0 , \quad \forall~i,j \qquad ; \qquad\frac{\partial \log P(\{\bm s\}|\{J, h\})}{\partial h_i}=0, \quad \forall~i.$$

Such procedure is termed \textit{Boltzmann machine learning}, because it relies on the equilibrium hypothesis, and it is characterized by the following steepest descent equations, cf. Eq. (\ref{GD}) for the updates of   the parameters $\Delta J_{ij}\equiv J_{ij}^{(t+1)}-J_{ij}^{(t)}$ and $\Delta h_{i}\equiv h_{i}^{(t+1)}-h_{i}^{(t)}$,

\begin{eqnarray}
\label{BM_C}
    \Delta J_{ij}&=&\eta \ \partial_{J_{ij}}\log P(\{\bm s\}|\{J, h\})  = \eta\ \beta\left[\bar C_{ij}   - \frac{1}{Z(\{J,h\})}\sum_{\{s\}} s_is_j\exp \{- \beta \mathcal{H}(\{s\}| \{J, h\})\}\right]  
    \nonumber
    \\ &=& \eta \ \beta \left[\bar C_{ij}   -\langle s_is_j\rangle\right],
    \\
      \Delta h_{i}&=&\eta \ \partial_{h_{i}}\log P(\{\bm s\}|\{J, h\}) = \eta \ \beta\left[\bar m_i   - \frac{1}{Z(\{J,h\})}\sum_{\{s\}} s_i\exp \{- \beta \mathcal{H}(\{s\}| \{J, h\})\}\right] 
      \nonumber
      \\
      &=& \eta\ \beta \left[\bar m_{i}   -\langle s_i\rangle\right],
      \label{BM_m}
\end{eqnarray}
where $$\langle s_i\rangle  = \frac{\partial \log Z}{\partial h_i},$$ is the theoretical model average magnetization, see Eq. (\ref{MEAN_M}),
     $$\langle s_i s_j\rangle  = \frac{\partial \log Z}{\partial J_{ij}}$$ is the theoretical two-spin disconnected correlation function, see Eq. (\ref{CORRELATION_C}), and $\eta$ parameter  \textit{learning rate} parameter mentioned in the introduction. It can be adjusted to optimize the convergence.
     One can then:
     \begin{enumerate}\item 
    start with trial values for $\{J,h\}$, 
    \item compute  the partition function $Z$, Eq. (\ref{PARTITION}), and the theoretical average magnetization and correlation, Eqs. (\ref{MEAN_M},\ref{CORRELATION_C}), at given $\{J,h\}$,
    \item compare with the empirical values (\ref{EXP_M}-\ref{EXP_C}) and compute the changes (\ref{BM_C},\ref{BM_m}) for $\{\Delta J,\Delta h\}$.
    \end{enumerate}
     Eventually, $\{J,h\}$ will converge to the parameter values maximizing the log-Likelihood. The downside of this procedure is that at each iteration step one has to compute $Z$ summing over all possible   configurations of $N$ spins.
     And they are $2^N$ in the Ising case, that is the simplest! 
     Their number grows with $N$ like $3^N$ for Blume Capel and it grows like $q^N$ for the clock model with $q$ hands, just to limit ourselves to variables taking a finite number of discrete values.
     This computation becomes quickly infeasible as $N$ becomes larger. An alternative is not using the exact partition function, but just an estimate. 
     For instance, $Z$ can be estimated on spin configurations generated by Monte Carlo simulations at equilibrium. This means that {\it at each step of the iterative Max-Likelihood (ML) procedure} new simulations have to be performed for up-to-date $\{J,h\}$ \cite{kappen1998efficient,tanaka1998mean,murphy2012machine}. Again, this procedure becomes costly as the size of the system increases.
     
To overcome these problems we can follow  approximate methods.
We will propose two approaches, the first one mimicking the max-likelihood method, but  employing the measurement of the whole set of spins, and not just their averages and correlations, Secs. \ref{Sec:MAX_PL},\ref{OVERFITTING},  \ref{LOGISTIC_REGRESSION} and the second one exploiting our knowledge of statistical physics, in particular of the analytic mean-field solution, that allows to have formulas for the $J$'s and $h$'s as functions of the averages and correlations of the sampled data, see Sec. \ref{INVERSE_MF}. 

In the first approach, termed max-\textit{Pseudo-}Likelihood, all spins, except one at a time, are taken from experimental configurations (assuming they are well thermalized). While this procedure, that we describe in detail in the next section, has no guarantee to be equivalent to the max-likelihood approach (except that in the operational unlikely limit of infinite number of $\{s\}$ measurements), it turns out to be quite effective and also seem to have applications outside standard statistical inference settings (for instance, in the context of associative memories \cite{dpseudo}).
In the second approach, termed {\it naive} mean-field, the simplest of a long series of more sophisticated mean-field models, a very direct inversion of the covariance matrix of the data infers the model coupling values.

\subsection{Max-Pseudo-Likelihood}
\label{Sec:MAX_PL}
In order to introduce the Pseudo-Likelihood (PL), we start from the Ising Hamiltonian, Eq. (\ref{eq:ising_ham}):
\begin{equation}
\label{HAM}
\mathcal H(\{s\},\{J,h\}) = -\sum_{i<j}^{1,N}J_{ij} s_is_j-\sum_{i=1}^Nh_is_i, 
\end{equation}
to which we have added (site-dependent) fields $h_i$. We rewrite it evidencing the contribution of a single spin $s_i$
\begin{eqnarray}
\nonumber
\mathcal{H}(\{s\}; \{J, h\}) &=& -\frac{1}{2}\sum_{ij}^{1,N}J_{ij}s_i s_j - \sum_{j=1}^{N}h_j s_j
\\
\nonumber
&=& -\sum_{j\neq i} J_{ij}s_i s_j - h_i s_i - \sum_{k<l} J_{kl}s_k s_l - \sum_{k\neq i} s_k h_k
\\
\label{HAM_REWRITTEN}
&=& \mathcal{H}_i(s_i|\{s_{\backslash i}\}) + \mathcal{H}_{\backslash i}(\{s_{\backslash i}\}),
\end{eqnarray}
where $\{s_{\backslash i}\}$ is the set of all spins excluding $i$. In the first line the sum runs over all indexes $i,j$. The partition function, Eq. (\ref{PARTITION}), can therefore be rewritten as 
\begin{equation*}
Z(\{J,h\}) = \sum_{\{s_1,\ldots,s_i,\ldots,s_N\}} \exp\{-\beta\mathcal{H}(\{s\}, \{J, h\})\}
\end{equation*}
\begin{equation*}
= \sum_{\{s_{\backslash i}\}} \exp\{-\beta\mathcal{H}_{\backslash i}(\{s_{\backslash i}\})\} \sum_{s_i=\pm 1} \exp\left\{ s_i \left( \beta \sum_{j \neq i} J_{ij} s_j + \beta h_i \right) \right\}
\end{equation*}
\begin{equation}
= \sum_{\{s_{\backslash i}\}} 2 \cosh \left( \beta \sum_{j \neq i} J_{ij} s_j + \beta h_i \right) \exp\{-\beta\mathcal{H}_{\backslash i}(\{s_{\backslash i}\})\}.
\end{equation}
Shortening the action on the isolated spin $s_i$ as
\begin{equation}
    \label{H_EFF}
    \tilde h(\{s_{\backslash i}\})= h_i+
    \sum_{j \neq i} J_{ij} s_j,
\end{equation}
we can rewrite the equilibrium distribution (\ref{DEF:BG_DIST}), dropping the dependence on $\{J,h\}$, in the equivalent form
\begin{equation}
    \label{BOLTZMANN_2}
    P(\{s\}) = P(s_i|\{s_{\backslash i}\})P(\{s_{\backslash i}\}) =  \frac{\exp\left\{-\beta s_i \tilde h(\{s_{\backslash i}\}) -\beta \mathcal{H}_{\backslash i}(\{s_{\backslash i}\})\right\}}
    { \sum_{\{s_{\backslash i}\}} 2 \cosh \left( \beta \tilde h(\{s_{\backslash i}\})\right) \exp\{-\beta\mathcal{H}_{\backslash i}(\{s_{\backslash i}\})\}}.
\end{equation}
In this reformulation the expectation values can be exactly rephrased as 
\begin{equation}
\langle s_i \rangle = \frac{1}{Z} \sum_{\{s_{\backslash i}\}} \exp \{-\beta \mathcal{H}_{\backslash i}(\{s_{\backslash i}\})\} \sum_{s_i=\pm 1} s_i \exp \left\{  \beta s_i\tilde h[s_{\backslash i}]  \right\}
\end{equation}
\begin{equation}
=
\frac{ \sum_{\{s_{\backslash i}\}}
2 \cosh \left( \beta \tilde h(\{s_{\backslash i}\})\right)  \exp \{-\beta \mathcal{H}_{\backslash i}(\{s_{\backslash i}\})\}  \tanh  \left( \beta \tilde h(\{s_{\backslash i}\})\right) }
{\sum_{\{s_{\backslash i}\}} 2 \cosh \left( \beta \tilde h(\{s_{\backslash i}\})\right) \exp\{-\beta\mathcal{H}_{\backslash i}(\{s_{\backslash i}\})\}}
\end{equation}
\begin{equation}
= \left\langle \tanh \left( \beta \sum_{j\neq i} J_{ij} s_j + \beta h_i \right) \right\rangle_{\mathcal{H}_{\backslash i}}
\end{equation}
for the mean magnetization and
\begin{equation*}
\langle s_i s_j \rangle = \frac{1}{Z} \sum_{\{s_{\backslash i}\}} s_j \exp \{-\beta \mathcal{H}_{\backslash i}(\{s_{\backslash i}\})\} \sum_{s_i=\pm 1} s_i \exp \left\{ \beta s_i  \tilde h(\{s_{\backslash i}\})\right\}
\end{equation*}
\begin{equation*}
= \frac{1}{Z} \sum_{\{s_{\backslash i}\}}  2 \cosh \left( \beta  \tilde h(\{s_{\backslash i}\}) \right) \exp \{-\beta \mathcal{H}_{\backslash i}(\{s_{\backslash i}\})\} \ s_j  \tanh \left( \beta \sum_{j\neq i} J_{ij} s_j + \beta h_i \right)
\end{equation*}
\begin{equation}
= \left\langle s_j \tanh \left( \beta \sum_{j\neq i} J_{ij} s_j + \beta h_i \right) \right\rangle_{\mathcal{H}_{\backslash i}}
\end{equation}
for the $2$-spins correlation.
Up to this point, we have manipulated exact identities. No approximations have been introduced, nor has any computational speedup been achieved.

At this point we can speed up the inference procedure by assuming that we can substitute a complete ensemble of configurations of $N-1$ spins with data samples,
\begin{equation}
    \label{pseudoAVE}
\langle \cdots \rangle_{\mathcal{H}_{\backslash i}} \simeq \frac{1}{M} \sum_{\mu=1}^{M} \cdots = \langle \cdots \rangle_M,
\end{equation}
such that, for a given a chosen spin, the canonical (Boltzmann-Gibbs) ensemble of the \(2^{N-1}\) configurations $\{s_{\backslash i}\}$ of the  $N-1$ spins other than $i$ is replaced by a sample of $M$ measured configurations.

Since the real statistical ensemble is sampled $M$ times and the total number of possible configurations is $2^{N-1}$, for $N$ even moderately large  one easily has \(M \ll 2^{N-1}\). 
The sampling is, thus, assumed to be statistically representative of the whole set of configurations. This is the fundamental hypothesis over which the max Pseudo-Likelihood approach is built.
Under our hypothesis this is necessary implied by the configurations being sampled at equilibrium, but simple equilibrium might not be sufficient.

\begin{figure}[t!]
\includegraphics[width=.95\textwidth]{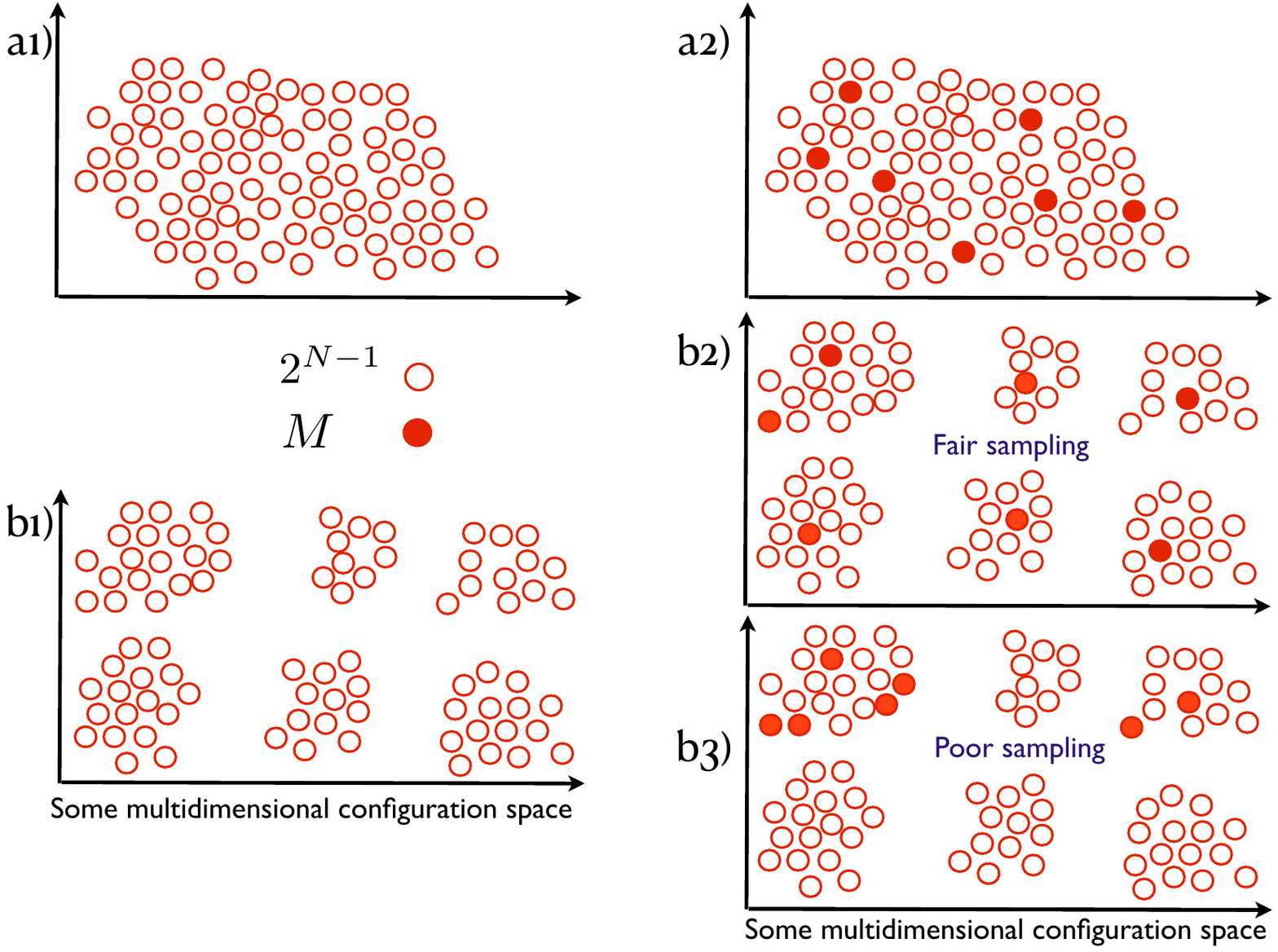}
\caption{\label{fig:sampling}Illustration of good and bad samplings, essential for the justification of the PL approximation. The open circles graphically  represent all possible system configurations in a two dimensional space. The full points are the measured configurations. In Fig. a1) the configurations are ideally points in an energy landscape in which no energy barrier separates them. In this case, no matter how the experimental points are distributed, they are typically a good sampling of the complete ensemble, see Fig. a2). In Fig. b1), instead, the configuration space is fragmented into different components. This is equivalent to an energy landscape in which different valleys are separated by barriers, i.e., there are more {\it equilibria}. In this case not all experimental samplings will provide a reasonable representation of the distribution of the system configurations. In Fig. b2) we sketch a possible fair representation, in which at least one configuration is sampled experimentally in each isolated component. In Fig. b3) this does not occur and only two components are sampled, heavily biasing the performance of the max PL inference procedure. When $M$ is not large, the second option is more likely to occur.}
\end{figure}

In Fig. \ref{fig:sampling} we give a pictorial sketch of two instances, one -- $a$ --   in which the $M$ configurations are representative in the convex multidimensional landscape in the $\{s\}$ variables (that is graphically reduced to a two dimensional space for typographical reasons) and one -- $b$ --  in which the $M$ experimentally measured configurations are not representative of the complete set of configurations, because of the lack of overall convexity in the energy landscape.  

In the framework of this approximation, starting from Eq. (\ref{BOLTZMANN_2}), we define the Pseudo-Likelihood \cite{Ravikumar2010} as the probability distribution of the spin  $s_i^{(\mu)} $ measured in the measurement $\mu=1,\ldots,M$ conditioned to the measured values $s_{\backslash i}^{(\mu)}$ of all spins but $i$, i.e. the $i$-th row PL:
\begin{equation}
    \label{P-PL}
P(\bm s_i|\bm s_{\backslash i}) = \prod_{\mu=1}^{M} \frac{\exp\left\{ -\beta \mathcal{H}_i(s_i^{(\mu)} | \{s_{\backslash i}\})^{(\mu)}]\right\}}{2\cosh\left(\sum_{j \neq i} \beta J_{ij} s_j^{(\mu)} + \beta h_i\right)},
\end{equation}
where we introduced the array in the measurement space of the spin $k$
\begin{equation}
{\bm s_k}= \{s_k^{(\mu)}\} = \{s_k^{(1)}, \ldots, s_k^{(M)}\}
\end{equation}
and, according to Eq. (\ref{HAM_REWRITTEN}),
\begin{equation}
\mathcal{H}_i(s_i|\{s_{\backslash i}\}) = -s_i \left( \sum_{j \neq i} J_{ij} s_j + h_i \right).
\end{equation}
We can, as previously done for the log-likelihood, define the single site $i$ \textit{log}-Pseudo-Likelihood, whose maximization is numerically easier to deal with:
\begin{equation}
    \label{iPL_ISING}
\mathcal{L}_i =\frac{1}{M} \log P(\bm s_i|\bm s_{\backslash i}) = \frac{1}{M}\sum_{\mu=1}^{M} \left[s_i^{(\mu)} \sum_{j \neq i}\beta  J_{ij} s_j^{(\mu)} +  \beta h_i s_i^{(\mu)} -  \log 2 \cosh \left( \sum_{j \neq i} \beta J_{ij} s_j^{(\mu)} + \beta h_i \right)\right].
\end{equation}
Here, the normalization $1/M$ is chosen to quantitatively compare PLs computed on datasets of different size $M$.

To maximize the Pseudo-Likelihood (\ref{iPL_ISING}) and find the most likely values of the model parameters  $\{J,h\}$, one can set up a iterative machine \footnote{$\mathrm{(}\star\mathrm{)}$ For those who studied Sec. \ref{MAX_LOG_LIKELIHOOD}, the procedure carried out solving Eqs. (\ref{GDPL_C},\ref{GDPL_m}) is similar  to the case of the max log-Likelihood case, see Eqs. (\ref{BM_C},\ref{BM_m}),
but with the exact averages now substituted by empirical averages}  in which the  {\it pseudo}-averages $\langle \ldots \rangle_M$ are prescribed by Eq. (\ref{pseudoAVE}):  
\begin{eqnarray}
\label{GDPL_C}
\Delta J_{ij}& =& \eta \ \beta \left[
\bar  C_{ij} - \left\langle s_j \tanh \left( \beta \sum_{j\neq i} J_{ij} s_j + \beta h_i \right) \right\rangle_M \right] \quad, \quad j\neq i
\\
\label{GDPL_m}
\Delta h_i &=& \eta \ \beta \left[
\bar m_i - \left\langle \tanh \left( \beta \sum_{j\neq i} J_{ij} s_j + \beta h_i \right) \right\rangle_M \right]
\end{eqnarray}
with $\bar m_i$ and $\bar C_{ij}$ the empirical magnetization and correlation over the spins, averaged over $M$ measured configurations:
\begin{equation}
\bar m_j = \frac{1}{M} \sum_{\mu=1}^M s_j^{(\mu)}
\end{equation}
\begin{equation}
\bar C_{ij} = \frac{1}{M} \sum_{\mu=1}^{M} s_i^{(\mu)} s_j^{(\mu)},
\end{equation}

Further details and considerations about the Max Pseudo-Likelihood inference method and its connection to machine learning are reported in the following Sections \ref{sss:IIIB1}-\ref{LOGISTIC_REGRESSION} 
$\mathrm{(}\star\mathrm{)}$.
Readers can skip these sections and move to Sec. \ref{INTERMEZZO} before facing the Mean-Field inference method in Sec. \ref{INVERSE_MF}

\subsubsection{$\mathrm{(}\star\mathrm{)}$ Imposing symmetric couplings }
\label{sss:IIIB1}
It is computationally very efficient to maximize a single site PL, cf. (\ref{iPL_ISING}), because all sites can be addressed in parallel. 
There is also a drawback in looking at a spin at a time, though. That is, from the max PL of $i$ we infer $J_{ij}$, whereas $J_{ji}$ is inferred by max $\mathcal L_j$. We know that theoretically $J_{ij}=J_{ji}$ \footnote{The whole procedure of Boltzmann machine learning is based upon the existence of a thermal equilibrium, in the canonical ensemble whose variables are distributed with the Boltzmann-Gibbs probability (\ref{DEF:BG_DIST}). If the couplings were not symmetric no equilibrium could be reached in the system dynamics and there would be no Hamiltonian. See e.g., Refs. \onlinecite{roudi2011mean, mezard2011exact} for the asymmetric Ising model.},
but maximizing  $\mathcal L_i$ and $\mathcal L_j$ in parallel does not guarantee that their estimators $J^{{\rm mpl}(i)}_{ij}$, $J^{{\rm mpl}(j)}_{ji}$ will be equal. Along a  heuristic analogy, then, 
usually a forced symmetrization is implemented as
$$J_{ij}=J_{ji}=\frac{J^{{\rm mpl}(i)}_{ij}+J^{{\rm mpl}(j)}_{ji}}{2}.$$
To  avoid the problem of asymmetric inferred couplings one can define an {\it all rows} Pseudo-Likelihood,
\begin{equation}
    \label{PL_ALL_ISING}
\mathcal{L} = \frac{1}{M}
\log \left[ P( \bm s_1|\bm s_{\backslash 1}) 
\times P(\bm s_2|\bm s_{\backslash 2}) \times \cdots \times P(\bm s_N|\bm s_{\backslash N}) \right] =  \frac{1}{M} \log  \prod_{i=1}^N P( \bm s_i|\bm s_{\backslash i})  = \sum_{i=1}^N \mathcal{L}_i, 
\end{equation}
in which all spins are considered as {\it pivot} spin once and each $J$ element is symmetric by construction (i.e., each $J_{ji}$ is written as $J_{ij}$, keeping, for instance, the convention of writing only the elements with $i<j$). Of course, this approach has the drawback of being numerically more demanding.

The PL approach, that we described in the case of the Ising model, can be easily generalized to more complicated models.

\subsubsection{$\mathrm{(}\star\mathrm{)}$ Pseudo-Likelihood for the Potts clock model }
In the case of the vectorial Potts model of Sec. \ref{POTTS} the Pseudo-Likelihood (\ref{iPL_ISING},\ref{PL_ALL_ISING}) generalizes as 
\begin{equation}
    \mathcal{L} = -\frac{1}{M} \sum_{i = 1}^N \sum_{\mu = 1}^M\left[ \log \sum_{c= 0}^{q-1} \exp \left \{\beta \sum_{j \neq i}J_{ij}\vec{s}_j^{\,(\mu)} \cdot (\vec{s}_c - \vec{s}_i^{\,(\mu)})\right \}\right], 
\end{equation}
where $\vec s_c=\{\sin \theta_c,\cos\theta_c\}$, cf. Eq. (\ref{THETA_POTTS}).

\subsubsection{$\mathrm{(}\star\mathrm{)}$ Pseudo-Likelihood for the Blume-Capel model }
In the case of the Blume-Capel model of Sec. \ref{BLUME}, in which $s_i=\{-1,0,1\}$, instead, the pseudo-log-Likelihood has the form:
\begin{equation}
    \mathcal{L} = \frac{1}{M}\sum_{i=1}^{N}\sum_{\nu=1}^{M} \left[\beta s_i^{(\nu)}\sum_{j\neq i}^{1,N} J_{ij}s_j^{(\nu)} - \beta \mu_i (s_i^{(\nu)})^2 - \log\left( 1+2e^{-\beta \mu_i}\cosh \sum_{j\neq i}^{1,N} \beta J_{ij}s_j^{(\nu)} \right)\right]
\end{equation}
A final observation is in order here. For the Pseudo-Likelihood to be convex it has to be \(M > \#\)coupling parameters.   If the function is not convex in the $\{J,h\}$ parameters, the maximisation does not provide a unique global solution and the max PL procedure is not reliable \cite{Aurell12}.

\subsubsection{$\mathrm{(}\star\mathrm{)}$ Overfitting and regularization}
\label{OVERFITTING}
An additional problem that may appear in inference is that, if too many parameters are involved in our learning, there is the danger of {\it overfitting}.
In this context, overfitting means, for instance, that the configuration of $\{J\}$ that we might learn by Pseudo-Likelihood has too many non-zero elements with respect to the real (unknown) network.
Since the ``machine'' performing the Pseudo-Likelihood maximization is trained over the experimental data, composing a so-called {\it training dataset}, the inferred couplings will probably be optimal for any data configuration belonging to the dataset. However, the coupling network thus reconstructed will not have a high value of Pseudo-Likelihood for any other measured spin configuration not included in  the original training set.  When this occurs it is said that the learning procedure does not generalize well. 
In classical statistical settings is often better to have less parameters to avoid overfitting and  provide a better generalization, in order to provide the optimal Boltzmann probability for any configuration \footnote{In modern machine learning settings the situation is more complicated. Indeed, overparametrization often helps Neural Networks to generalize well. The question on why this is the case is still open \cite{oneto2023we}.}.
 
The strategy that we will follow here to prevent overfitting is {\it regularization}.
We need for a regularizer item forcing irrelevant couplings to be zero. In order to build one reasonable regularization we introduce the $\ell_n$-norm $||x||_n$ of a generic set of parameters $x$:
$$
||x||_n = \sqrt[n]{|x_1|^n + \cdots + |x_N|^n}.
$$
In particular we will use the $\ell_1$ and the $\ell_2$ norms in introducing the $\ell_1$- and the $\ell_2$-regularized $\{J,h\}$ optimal parameter set maximizing the ($i$-)Pseudo-Likelihood:
$$
\{J_{ij}^{\inf}, h_i^{\inf}\}^{\ell_1}_{j \in \partial i} = \underset{\{J_{i \partial i}, h_i\}}{\argmax}\left[ \mathcal{L}_i (\{J_{i \partial i}, h_i\}) + \lambda_J \sum_{j\neq i} |J_{ij}| + \lambda_h h_i \right],
$$
$$
\{J_{ij}^{\inf}, h_i^{\inf}\}^{\ell_2}_{j \in \partial i} =\underset{\{J_{i \partial i}, h_i\}}{\argmax} \left[ \mathcal{L}_i (\{J_{i \partial i}, h_i\}) + \lambda_J \sum_{j\neq i} J_{ij}^2+ \lambda_h h_i^2 \right],
$$
where $\partial i$ is the set of spins neighboring the spin $i$.
The regularizers $\lambda_J$ and $\lambda_h$ must not be too large, in order not to change the Pseudo-Likelihood too much, and underestimating couplings and field too much. At the same time, they should not be too small either, otherwise their effect is negligible and overfitting remains.
There are systematic methods to determine reasonable values of the regularizers, like {\it cross-validation} \cite{fisher2014}. See, e.g., Ref. \onlinecite{Marruzzo2018} for an instance of such a method combined with max PL.
In practice, it is often faster to find some robust interval of values for the $\lambda$'s by basic trial and error. Though apparently more akin to a heuristic parameter selection than to statistical inference, this is often a valid way.

The $\ell_1$ regularization (based on the sum of absolute values of the parameters) is often called {\it Least Absolute Shrinkage and Selection Operator}, i.e., lasso \cite{Hastie2015}. It is the more drastic regularization of the $\ell_n$ family, setting to zero many $J$'s and, thus, being prone to inferring {\it sparse} networks.
By sparse it is meant that the number of non-zero interactions of a spin with the others does not grow with the number of spins in the system.
Therefore, lasso is very good when the underlying graph is actually sparse. 

The $\ell_2$ is a softer regularization, and tends to reduce but not eliminate small couplings. That is good if the network is not sparse, but dense (eventually a {\it complete graph} in which  each spin is connected to all the others. This is why a complete graph is often refereed to as a fully connected graph).  Problems come about if the original network is sparse, because $\ell_2$ will tend to infer a small (maybe even extremely small), but larger than zero, value even for non-existing connections. 

On the contrary, much evidence has been collected \cite{Hastie2015} that lasso is able to perform well also on fully connected networks. Indeed, provided the regularizer is well tuned,  this kind of graphs are accurately reconstructed also by means of a $\ell_1$ regularization. This is the origin of the quote ``bet on sparsity'':{\it it is better to use a procedure that does well in sparse problems, since no procedure does well in dense problems} \cite{Hastie2015}. Stretching the sentence a bit, this is akin to saying that, even if you are wrong about the model connectivity, you can still infer the right network parameters.

A further consideration regarding regularization from the point of view of Bayesian inference. 
If we consider (Pseudo-)Likelihood rather than log-(Pseudo-)Likelihood, we realize that setting a regularization to the Likelihood amounts to choosing  a non-uniform prior in Eq. (\ref{BAYES_INFERENCE}). In particular, this corresponds to choosing a prior distribution of the exponential of the absolute value  for the $J$ in the $\ell_1$ case and a Gaussian distribution of zero mean in the $\ell_2$ case.

Several other approaches can be adopted to reduce the number of relevant parameter to be learned, in the spirit of lasso. We can very synthetically refer to them as {\it information criteria} (IC), that is, criteria by which one can decide where the most of information lies and disregard the rest. Instances of such criteria are functions of the number of inferred parameters, also depending  on data size $M$, such as the Akaike IC \cite{Akaike74}, the Bayesian IC \cite{Schwarz78,Kass95} or the Decimation  IC \cite{Decelle14,yamanaka_15}. Progressively reducing the number of parameters (or progressively increasing them starting from a {\it tabula rasa} scenario \cite{Engel01}), the various IC functions reach a min (Akaike, Bayes) or a max (Decimation) in correspondence to the best estimation of the number of parameters of the true model. In our case, this number corresponds to the number of non-zero couplings in the Ising spin graph.

\subsubsection{{\color{black}{$\mathrm{(}\star\mathrm{)}$ Logistic regression and max pseudo-likelihood}} }
\label{LOGISTIC_REGRESSION}
We  highlight that 
the probability distribution of a single measure of a spin $s_i$ in the Pseudo-Likelihood approximation can be rewritten, see Eq.   (\ref{P-PL}), as a {\it logistic regression} function. Indeed, 
\begin{equation}
    \label{P-PL-GL}
P(s_i|s_{\backslash i}) =  \frac{\exp\left\{ -\beta s_i \tilde h_i(\{s_{\backslash i}\})\right\}}{2\cosh\left(\beta \tilde h_i(\{s_{\backslash i}\})\right)}  = \frac{1}{1+\exp\left\{-2 \beta s_i \tilde h_i(\{s_{\backslash i}\})\right\}} = \frac{1}{1+e^{-z_i}},
\end{equation}
where we introduce the {\it decision boundary} variable $z_i = 2 \beta s_i \tilde h_i(\{s_{\backslash i}\})$ to show how the Pseudo-Likelihood function is a sigmoid and its maximization in $\{J,h\}$ corresponds to a logistic regression problem.

\subsubsection{An intermezzo on energy landscape of the  direct space and temperature noise in data collection }
\label{INTERMEZZO}

In the Boltzmann-Gibbs probability distribution, Eq. (\ref{DEF:BG_DIST}), we have always kept the model parameters we wanted to infer and the inverse temperature $\beta$ separate. However, these parameters and $\beta$ always appear as a product, so that in order to infer the true parameter values, such as the $J$, one has to divide by $\beta$ after the inference has been carried out. 
Indeed, the temperature is redundant as a free parameter in the inference problem (not in sampling, as we will see). 
Its effect is to rescale the couplings $J_{ij}$ by a factor $\beta$, 
$\beta J_{ij}\to J_{ij}$, thus strengthening them in the low-temperature regime. 
Technically it is enough to set $\beta=1$, e.g., in  Eqs. (\ref{GDPL_C},\ref{GDPL_m}) or $\mathrm{(}\star\mathrm{)}$ in Eqs. (\ref{BM_C},\ref{BM_m}).

Things are more complicated for data generation, because temperature can have a very important role in the generation of the data that make the inference possible. In particular, this temperature dependence becomes relevant when phase transitions happen and, therefore, the low-temperature phase is very different from the high-temperature one. We will now describe these phenomena in more detail.

\begin{figure}[t!]
\includegraphics[width=.8\textwidth]{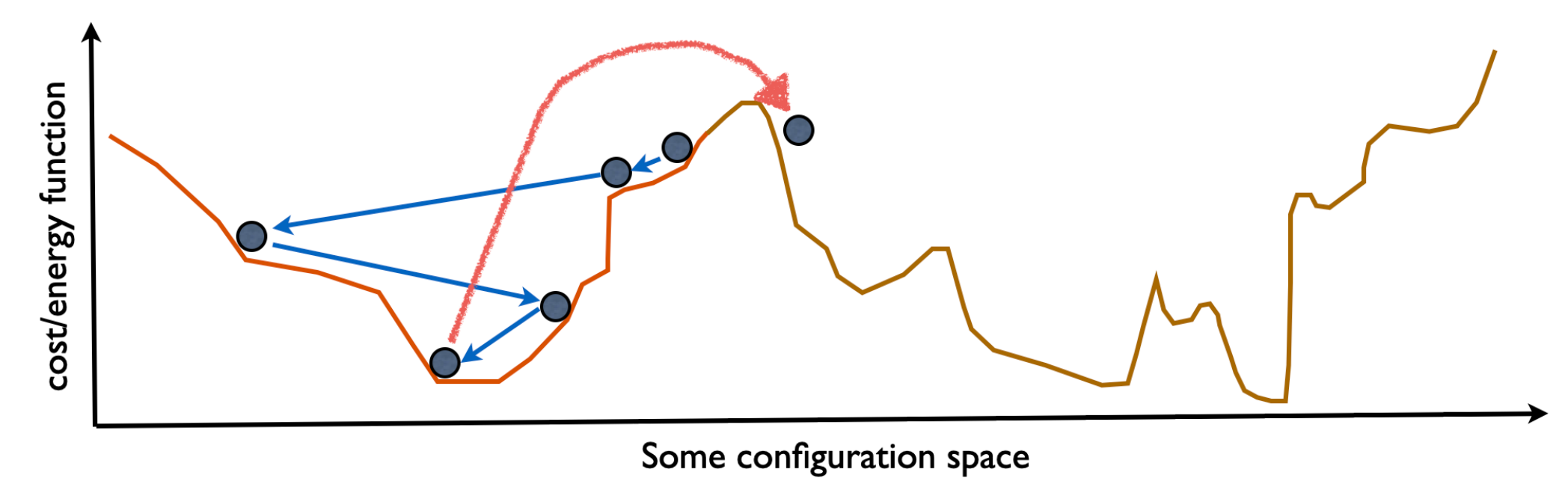}
\caption{\label{fig:landscape} Illustrative sketch of an energy function of the configurations of data in which more minima, both global and excited, are present. A gradient descent (blue arrows) ends up stuck in the first minimum it reaches, which depends only on the initial conditions. A sampling at non-zero temperature (red arrow), on the other hand, allows also to overcome barriers and explore more minima in the energy landscape, thus allowing for better sampling the  configurational space of the data to be fed to the learning procedure.}
\end{figure}

In Fig. \ref{fig:sampling}, when  we took into consideration the assumptions behind the max PL method, we pictorially displayed two cases mentioning a landscape in the energy (or cost, or loss) function of the variables configurations. In Fig. \ref{fig:landscape} we illustrate a one dimensional reduction of an energy landscape in which, given some fixed $\{J,h\}$ configuration, more minima are present.
Data can be sampled more or less properly, in this case, depending on how many configurations in the dataset pertain to each minimum, representing a state of the system. If no measured configuration belongs to a given state, any inference method will provide biased results. This holds, for instance, both for the max PL and the mean-field methods (the latter is based on the inversion of the two-spin correlation function matrix that we will introduce in the next section \ref{INVERSE_MF}).
Sampling data at zero temperature is equivalent to a gradient descent in the variable space, the $\{s\}$. 
In a complex landscape it will often result in a sampling of configurations stuck at relatively high energy, even in very shallow minima, and rarely reaching deeper global minima, representing the ground states. In the learning of the $\{J,h\}$ from the data in the inverse problem these biased dataset will cause the impossibility of inferring a network similar to the original one, or, even, any plausible network. Indeed, on data acquired at very low $T$ the inference procedure might not converge at all. 

This is why the introduction of a temperature in the sampling process, that is, a stochastic source in the gradient descent that allows also for upward jumps in the relaxation dynamics, will favor the probing of different ground states.
A too large temperature, of course, will only produce completely random $\{s\}$ configurations, for which the mutual interaction or any effect of external fields is irrelevant and the data will not have enough structure to distinguish optimal couplings and fields.
In section \ref{Sec:SI} we will see the role of temperature in different statistical mechanics models.

\subsection{Mean-Field}
\label{INVERSE_MF}
Another method, of relatively simple application, to infer the couplings and fields of a statistical mechanics  model starting from measurements of the spins is based on the  mean-field theory developed to study the thermodynamic behavior of such models in high dimension (e.g., in $D\geq 4$ for the nearest-neighbor ferromagnetic Ising model) or any model on sparse random graphs, for which no underlying geometry is defined for the connectivity. 
This method is exact in those cases. It is, on the other hand, just an approximation for Ising, Potts or Blume-Capel models in ``low dimension", i.e., below some {\it upper critical dimension} $D<D_{\rm upc}$ that depends on the model connection network  and values. Despite this, it is still qualitatively valid because it predicts the phase transition occurring in these models. For the ferromagnetic Ising model, for instance, the mean-field theory predicts the existence of a phase transition, which actually occurs in the real system in $D=2,3$ (not in $D=1$, though),  although with a critical behavior quantitatively different from the  mean-field one.

The method is easy to apply and does not even need the whole set of measured configuration $\{\bm s\}$, but just its average vector, of components
$m_i=\langle s_i\rangle $, cf. Eq. (\ref{MEAN_M}) and the two-point connected correlation matrix (also called  covariance). 
In particular, the experimental covariance $\Gamma_{ij}$, cf. Eqs (\ref{EXP_C},\ref{EXP_M}),  reads
\begin{equation}
\label{EXP_COVA}
\Gamma_{ij} = \frac{1}{M} \sum_{\mu=1}^{M} s_i^{(\mu)} s_j^{(\mu)} - \frac{1}{M} \sum_{\mu=1}^{M} s_i^{(\mu)} \frac{1}{M} \sum_{\mu=1}^{M}s_j^{(\mu)}.
\end{equation}
We anticipate that in the simplest case (the so-called {\it naive} mean-field) the final formula to infer the network couplings $J^{\rm MF}$ is simply
\begin{equation}
 \beta J^{\rm MF}_{ij} = - (\Gamma^{-1})_{ij}\quad (i \neq j).
 \label{MF:J_C}
\end{equation}

The derivation of Eq. \eqref{MF:J_C} requires a bit of knowledge of the mean-field theory of critical phenomena and of variational principles in statistical physics. For the readers who are interested, in the next  sections \ref{Sec:MF_dir}-\ref{sss:IIIC5} $\mathrm{(}\star\mathrm{)}$  we try to give a simple presentation to the ``ingredients'' we need. For a broader and deeper analysis  readers may further refer to Refs. \onlinecite{Tanaka2000,Opper2001,nguyen2017inverse}.

The results obtained using  formula \eqref{MF:J_C} are  reported in Sec. \ref{Sec:SI} and compared to the outcome of the Max Pseudo-Likelihood method. Who is interested in its application can, thus, directly skip the rest of Section III.C and move to Section  \ref{Sec:SI}.

\subsubsection{$\mathrm{(}\star\mathrm{)}$ Mean-field theory for the direct problem }
\label{Sec:MF_dir}
In this approximation the influence of the thermal fluctuations of a variable (spin) on the other coupled variables  are neglected, and vice-versa. In formulas, this amount to have vanishing connected correlation functions, Eq. (\ref{COVARIANCE_G}), in the large $N$ limit:
\begin{equation}
\label{clustering}    
\Gamma_{ij}=\langle s_is_j\rangle-\langle s_i\rangle\langle s_j\rangle\underset{N\to\infty}{\simeq} 0, \; \; \; \; \; i \neq j,
\end{equation}
where the average is taken over the equilibrium distribution Eq. (\ref{DEF:BG_DIST}).
This is automatically satisfied if we approximate $s_is_j\simeq s_im_j+m_is_j-m_im_j$, being $m_k=\langle s_k\rangle$. Hamiltonian (\ref{HAM}) therefore becomes

\begin{equation}
\label{HAM_MF}
\mathcal H_{\rm MF}(\{s\},\{J,h\}) = -\sum_{i<j}^{1,N}J_{ij} s_is_j-\sum_{i=1}^Nh_is_i \simeq -\sum_{i=1}^N\tilde{h}_i(\{m\})s_i
+\frac{1}{2}\sum_{ij}^{1,N}J_{ij} m_i m_j, 
\end{equation}
where we defined
the {\it mean-field} on spin $i$ as
\begin{equation}
    \label{MEAN-FIELD}
    \tilde{h}_i(\{m\}) = \sum_{j=1}^N
    J_{ij}m_j + h_i.
\end{equation}
We can now compute the equilibrium mean-field probability distribution, starting from its normalization, the mean field partition function
\begin{eqnarray}
    \label{MEAN-FIELD-Z}
    Z_{\rm MF} &=& \sum_{\{s\}}e^{-\beta \mathcal H_{\rm MF}(\{s\},\{J,h\})}
    =\exp\left\{-
\frac{\beta}{2}\sum_{ij}^{1,N}J_{ij} m_i m_j\right\}
 \sum_{\{s\}}e^{\beta\sum_{i=1}^N\tilde{h}_i(\{m\})s_i}
 \\
 \nonumber
 &=&
 A(\{m\})\prod_{i=1}^N \left[2 \cosh\left(\beta \tilde{h}_i(\{m\})s_i\right)\right]
 =\prod_{i=1}^N z_{\rm MF}^{(i)},
 \\
 z_{\rm MF}^{(i)} &\equiv&  2 A(\{m\})\cosh\left(\beta \tilde{h}_i(\{m\})s_i\right),
 \\
  A(\{m\}) & \equiv&\exp\left\{-
\frac{\beta}{2}\sum_{ij}^{1,N}J_{ij} m_i m_j\right\}.
\end{eqnarray}
We can observe that it is factorized in single site partition functions $z^{(i)}$.
Also the numerator of the Boltzmann-Gibbs distribution is, of course, factorized in the mean-field case:
$$e^{-\beta \mathcal H_{\rm MF}(\{s\},\{J,h\})}=A(\{m\})\prod_{i=1}^N
e^{\beta\tilde{h}_i(\{m\})s_i}$$
and, therefore,
\begin{eqnarray}
    \label{BG-MF}
    P_{\rm MF}(\{s\}) &=& \frac{e^{-\beta \mathcal H_{\rm MF}(\{s\},\{J,h\})}}{Z_{\rm MF}} = \prod_{i=1}^N p_i(s_i),
    \\
    \label{BG-MF-FACTOR}
   p_i(s) &=&\frac{e^{\beta\tilde{h}_i(\{m\})s}}{2 \cosh\left(\beta \tilde{h}_i(\{m\})s\right)},
\end{eqnarray}
that is, the joint distribution  $P_{\rm MF}(\{s\})$ of a spin configuration $\{s\}$ factorizes into  single spin distributions. This is, actually, a necessary and sufficient condition for mean-field. The approximation can, indeed, be defined just imposing the factorization of the joint probability distribution of the interacting variables, rather then Eq. (\ref{clustering}).
We may rigorously define the mean-field approximation imposing this condition in the first place. 
The most generic distribution for a Ising-like variable $s=\pm 1$ reads
\begin{eqnarray}
    p_i(s)=\pi_i\delta(s-1)+(1-\pi_i)\delta(s+1)
    \label{SPIN-DISTRIB}
\end{eqnarray}
and we just have to determine the weight $\pi$, e.g., as a function of the average magnetization $m_i$. Indeed,
$$m_i=\sum_{s=\pm 1} s\  p_i(s)= \pi_i - (1-\pi_i)$$
and, therefore,
\begin{eqnarray}
    \pi_i=\frac{1+m_i}{2}.
    \label{SPIN-WEIGHT}
\end{eqnarray}
Combining Eq. (\ref{SPIN-DISTRIB}) and Eq. (\ref{SPIN-WEIGHT}) the single spin distribution is equivalent to Eq. (\ref{BG-MF-FACTOR}).
The reader can easily prove it, once the mean-field average magnetization is computed:
\begin{eqnarray}
\label{MF_EQS}
    m_i^{\rm mf}=\tanh\left(\beta \tilde h_i(\{m^{\rm mf}\})\right) \qquad i=1,\ldots,N .
\end{eqnarray}
This is the {\it naive} mean-field set of equations. It is 
naive because it does not include the so-called \textit{Onsager reaction term}, which is instead required for a more sophisticated approximation. In order to deal with heterogeneous, disordered systems, one needs to go beyond this approximation, as did Thouless, Anderson and Palmer in their fundamental paper Ref. \onlinecite{Thouless77}. We will not address the issue here, but the interested reader can refer to  Refs. \onlinecite{tanaka1998mean,kappen1998efficient}.  

\subsubsection{$\mathrm{(}\star\mathrm{)}$ Helmholtz and Gibbs free energies }
 Before dealing with variational principles of the Helmholtz and Gibbs free energies we recall their definition and use.
Let us start from the Helmholtz free energy, whose equilibrium definition is
\begin{eqnarray}
    \label{FREEN}
    F(\{J,h\})=-\frac{1}{\beta}\ln Z(\{J,h\}).
\end{eqnarray}
We will set $\beta=1$ from now on, incorporating $\beta$ in $\beta J_{ij}$ or $\beta h$. Indeed, as we mentioned in Sec. \ref{INTERMEZZO}, when we are inferring the parameters from the data under the Boltzmann-Gibbs distribution assumption the temperature is just a rescaling.
Its Legendre transform is
the Gibbs free energy \cite{Touchette2009,nguyen2017inverse,Leuzzi2025}
\begin{eqnarray}
    \label{FREEN_GIBBS}
    G(\{J\},\{m\})= \underset{\{h\}}{\max}
    \left\{
    \sum_{i=1}^N h_i m_i +  F(\{J,h\})
    \right\},
\end{eqnarray}
where the max condition provides the relationship between the conjugated variables $\{h\}$ and $\{m\}$ variables
\begin{eqnarray}
    m_i=-\frac{\partial F}{\partial h_i}, \qquad \forall~i \qquad \rightarrow \qquad  h_i=h_i(\{m\}).
\end{eqnarray}
Since the Legendre transform is involutive, its Legendre transform is the Helmholtz free energy itself, that is 
\begin{eqnarray}
    \label{FREEN_HELM}
   F(\{J,h\}) = \underset{\{m\}}{\max}
    \left\{
  -  \sum_{i=1}^N h_i m_i +  G(\{J\},\{m\})
    \right\}
\end{eqnarray}
with 
\begin{eqnarray}
    h_i=\frac{\partial G}{\partial m_i}, \qquad \forall~i \qquad \rightarrow \qquad  m_i=m_i(\{h\})
    \label{h_i_GIBBS}.
\end{eqnarray}
The latter Eq. (\ref{h_i_GIBBS}) will be fundamental in the inverse problem.
Another cornerstone formula is the second derivative of $G$:
\begin{eqnarray}
    \frac{\partial^2 G}{\partial m_i\partial m_j}&=&\frac{\partial h_i}{\partial m_j} 
    \label{IC_ij_GIBBS}.
\end{eqnarray}
 The right hand side can be obtained by its inverse
$$ \frac{\partial m_i}{\partial h_j} = -\frac{\partial^2 F}{\partial h_i \partial h_j } = \Gamma_{ij},$$
that can be explicitly computed using Eqs. (\ref{FREEN}), (\ref{PARTITION}), with $\beta=1$. 
Now, the matrix of elements $$\frac{\partial h_i}{\partial m_j} $$ is the Jacobian matrix of the transformation $\{m \}\to \{h \}$  and the theorem of inverse functions guarantees that 
$$\left(\frac{\partial h}{\partial mj}\right)_{ij}=\left(\frac{\partial m}{\partial h}\right)^{-1}_{ij}=\left(\Gamma^{-1}\right)_{ij}.$$
Eventually one has
\begin{eqnarray}
    \frac{\partial^2 G}{\partial m_i\partial m_j}&=&\left(\Gamma^{-1}\right)_{ij}
\end{eqnarray}
to be noted for later use.

\subsubsection{$\mathrm{(}\star\mathrm{)}$ Free energies variational principles }
\label{Sec:VARIATIONAL}
The second fundamental kind of ingredient to yield the inference mean-field formula (\ref{MF:J_C}) are the variational principles of the free energies. That is, a version of the free energies in which the distribution $P(\{s\})$ is not the Boltzmann-Gibbs distribution with the right values of $\{J,h\}$
but some unknown distribution $Q(\{s\})$.
In a generic form, therefore, the Helmholtz free energy functional of the distribution $Q$ can be written as (once again $\beta=1=T$)
$$F[Q]=U[Q]-S[Q],$$
with the variational internal energy and entropy given by
\begin{eqnarray}
    U[Q]&=& \langle \mathcal H(\{s\})\rangle_Q = -\sum_{i=1}^N
h_i\langle s_i\rangle_Q -\sum_{i<j}^{1,N}J_{ij} \langle s_is_j\rangle_Q  
\label{VAR_U}\\
\label{VAR_S}
    S[Q]&=& -\langle \ln Q(\{s\})\rangle_Q.
\end{eqnarray}
The notation $\langle\ldots\rangle_Q$ denotes the average over distribution $Q$.
One can prove that, by varying $Q$ in order to find the minimum $F[Q]$, one finds the equilibrium Helmholtz free energy $F(\{J,h\})$:
$$\underset{\{Q\}}{\min} \ F[Q]=F(\{J,h\}).$$
For what concerns its Legendre transform
the variational form reads
\begin{eqnarray}
    G[Q] &=& \underset{\{h\}}{\max}\left\{
    \sum_{i=1}^N h_i m_i +F[Q]
    \right\}
    \\
    \nonumber
    &=&\underset{\{h\}}{\max}\left\{
    \sum_{i=1}^N h_i \left(m_i-\langle s_i\rangle_Q\right) -\sum_{i<j}^{1,N}J_{ij} \langle s_is_j\rangle_Q -S[Q]
    \right\}.
\end{eqnarray}
Also in the Gibbs case the equilibrium free energy $G(\{J,m\})$ is realized by the $Q$ distribution minimizing the functional $G[Q]$. Moreover,  
 we can restrict ourselves to look only at distributions $Q$ such that the average magnetizations coincide with the actual local average magnetizations $m_i$, in which case 
 \begin{eqnarray}
     \label{GIBBS_MIN_RESTR}
     G(\{J,m\})=\underset{\{ Q|\langle \vec s\rangle_Q = \vec m\}}{\min} G[Q]= \underset{\{ Q|\langle \vec s\rangle_Q = \vec m\}}{\min}\left\{-\sum_{i<j}^{1,N}J_{ij} \langle s_is_j\rangle_Q -S[Q]
     \right\}.
 \end{eqnarray}

\subsubsection{$\mathrm{(}\star\mathrm{)}$ Inverse naive mean-field statistical inference }
\label{Sec:NMF}
Let us  put together the mean-field approximation and the variational principle approach by introducing  factorized probability distributions for the spin configurations:
\begin{eqnarray}
    Q_{\rm mf}(\{s\})=\prod_{i=1}^Nq_i(s_i),
\end{eqnarray}
such that $\langle s_i\rangle_Q = \langle s_i\rangle_{q_i}=\bar s_i$, $\forall~i$, the latter being the empirical average on data of the local magnetization, Eq. (\ref{EXP_M}). Observing that the most general form of a function of a Ising-like variable is $q_i(s)=A_i+B_i s$, with the constraints
\begin{eqnarray}
    \sum_{s=\pm 1} q_i(s)=1,
    \label{CLOSURE_MF}
    \\
    \sum_{s=\pm 1} s\  q_i(s) = \tilde m_i,
\end{eqnarray}
we find 
\begin{equation}
    q_i(s)=\frac{1+\tilde m_i}{2}.
    \label{q_MF}
\end{equation}
Because of factorization two-point correlation functions are factorized too, $\langle s_i s_j\rangle_Q= \tilde m_i \tilde m_j$, and the connected correlation functions are null, coherently with the initial mean-field approximation definition, Eq. (\ref{clustering}). 
 In the variational Helmholtz
free energy, then, the internal energy functional (\ref{VAR_U})  take the form
\begin{eqnarray}
    U[Q_{\rm mf}]=   -\sum_{i=1}^N
h_i\tilde m_i -\sum_{i<j}^{1,N}J_{ij} \tilde m_i \tilde m_j =  U_{\rm mf}(\{\tilde m\}),
\label{VAR_U_MF}
\end{eqnarray}
and, according to Shannon, the entropy functional (\ref{VAR_S}) reads
\begin{eqnarray}
    \nonumber
    S[Q_{\rm mf}]&=& - \sum_{\{s\}} Q_{\rm mf}(\{s\})\log Q_{\rm mf}(\{s\}) 
    = - \prod_{i=1}^N \sum_{s_i=\pm 1}
   \prod_{i=1}^N  q_i(s_i)\log \left(
  \prod_{j=1}^Nq_j(s_j)  \right)
  \\
  \nonumber
  &=& - \sum _{j=1}^N \prod_{i=1}^N\sum_{s_i=\pm 1} q_i(s_i)\log 
  q_j(s_j)  = - \sum _{j=1}^N \sum_{s_j=\pm 1} q_j(s_j)\log 
  q_j(s_j) 
  \\
  \label{VAR_S_MF}
  &=& - \sum _{j=1}^N\left[\frac{1+\tilde m_i}{2}\log \frac{1+\tilde m_i}{2}+ \frac{1-\tilde m_i}{2}\log \frac{1-\tilde m_i}{2}
  \right]= S_{\rm mf}(\{\tilde m\}),
\end{eqnarray}
where we used Eqs. (\ref{CLOSURE_MF}), (\ref{q_MF}).
The variational principle, thus, reduces to
\begin{eqnarray}
\nonumber
    \underset{\{Q\}}{\min} \ F[Q]&=&F(\{J,h\}) =\underset{\{Q_{\rm mf}\}}{\min} \ F[Q]=F(\{J,h\})\\ \nonumber
&=&    \underset{\{\tilde m\}}{\min} \ F_{\rm mf}(\{\tilde m\})=  \underset{\{\tilde m\}}{\min} \ \left[U_{\rm mf}(\{\tilde m\})- S_{\rm mf}(\{\tilde m\})\right].
    \end{eqnarray}

 We only need to derive $U_{\rm mf}$ and $S_{\rm mf}$ with respect to the parameters $\{\tilde m\}$:
 \begin{eqnarray}
     \frac{\partial U_{\rm mf}}{\partial \tilde m_i} &=&- h_i-\sum_jJ_{ij}\tilde{m}_j= - \tilde h_i(\{\tilde{m}\})
     \\
      \frac{\partial S_{\rm mf}}{\partial \tilde m_i} &=&-\frac{1}{2}\log \frac{1+\tilde m_i}{1-\tilde m_i} = - \atanh \tilde{m}_i
 \end{eqnarray}
in order to see that the minimum variational free energy is obtained for parameters $\tilde m$ satisfying the system equation
 \begin{equation}
     \tilde m_i = \tanh \tilde h_i(\{\tilde{m}\}), \qquad \forall~i=1,\ldots,N,
 \end{equation}
 i.e., Eq. (\ref{MF_EQS}) for the (naive) mean-field local magnetizations. We can, thus, drop the tilde.

 Our objective is to infer $\{J,h\}$ parameters.
  To estimate the $\{J\}$ we use the principle of minimum Gibbs free energy, applied to mean field distributions:
  \begin{eqnarray}
   G_{\rm mf}(\{J,m\})&=&\underset{\{ Q_{\rm mf}=\prod_i q_i|\langle  s_i\rangle_{q_i} = m_i\}}{\min}
   \left\{-\sum_{i<j}^{1,N}J_{ij} \langle s_is_j\rangle_{Q_{\rm mf}} -S[Q_{\rm mf}]\right\}
   \\
   &=&-\sum_{i<j}^{1,N}J_{ij} m_i m_j -S_{\rm mf}(\{m\})
  \end{eqnarray}
Deriving once with respect to $m_i$ we have, using Eq. (\ref{h_i_GIBBS}),
\begin{equation}
     h_i=\frac{\partial G_{\rm mf}(\{J,m\})}{\partial m_i} = -\sum_{j=1}^N J_{ij} m_j + \atanh m_i,
\end{equation}
and deriving twice, using Eq. (\ref{IC_ij_GIBBS}),
\begin{equation}
     \left(\Gamma^{-1}\right)_{ij}=\frac{\partial^2 G_{\rm mf}(\{J,m\})}{\partial m_i\partial m_j} = -J_{ij}, \qquad i\neq j,
\end{equation}
as anticipated in Eq. (\ref{MF:J_C}).
Inserting the experimental averages (\ref{EXP_M},\ref{EXP_COVA}) for $m$'s and $\Gamma$'s in the above formulas we obtain a direct estimate of the coupling constants and external magnetic field of the Ising model under probe.

\subsubsection{$\mathrm{(}\star\mathrm{)}$ Inverse mean-field for the vectorial Potts and the Blume-Capel models }
\label{sss:IIIC5}
In the case of the vectorial Potts model, cf. Sec. \ref{POTTS} we have two covariance matrices, $\mathbf{\Gamma_x}$ and $\mathbf{\Gamma_y}$, corresponding to the $x$ and $y$ components of the spins. Then the best estimate of the couplings is given by
\begin{equation}
   - J_{ij} = \frac{1}{2}[(\mathbf{\Gamma_x})^{-1}_{ij} + (\mathbf{\Gamma_y})^{-1}_{ij}].
\end{equation}
Notice that using just one covariance matrix, as would be done in the Ising case, results in a loss of information. 

In the case of the Blume-Capel model, see Sec. \ref{BLUME}, the coupling matrix is still given by the inverse of the mean-field covariance, as in the Ising model.

\section{Statistical inference on interacting statistical physics systems}
\label{Sec:SI}
Once the methodology has been exposed and the basic formulas for the network and fields reconstruction have been derived, we now move on to specific estimates on different models and underlying graphs. We test the procedures on known models using synthetically generated data. A couple of simple key tools that we will use to analyze the quality of the inference are the reconstruction error and the ranking plot. We first look at the reconstruction error properties of  on all the tested models with the Max Pseudo-Likelihood and the Mean-Field techniques introduced in Secs. \ref{Sec:MAX_PL}, \ref{INVERSE_MF}.
In the following section we display the ranking plots and compare the couplings sorted by magnitude.  

\subsection{Reconstruction error}
As a first test, to check the validity of the reconstructions,  we introduce the reconstruction error $\gamma_J$  to compare the original $\{J\}$ and the inferred $\{\hat J\}$ system couplings:
\begin{equation}
    \gamma_{J} = \sqrt{\frac{\sum_{ij}(\hat{J}_{ij}-J_{ij})^2}{\sum_{ij}(J_{ij})^2}}.
\end{equation}
where $J_{ij}$ are the original couplings and $\hat{J}_{ij}$ are the reconstructed ones.

As we mentioned in Sec. \ref{INTERMEZZO} data might be acquired in presence of thermal noise. Knowing the temperature (and we do since we are generating our test data) we can observe the role of the original system thermal noise in the inverse problem of reconstructing the network couplings.
We, thus, hereafter present the $\gamma_J$ behavior as a function of the data temperature $T$ in the Ising model, as well as, $\mathrm{(}\star\mathrm{)}$ in the vectorial Potts and the Blume-Capel models. 

\subsubsection{The Ising Model}

\begin{figure}[t!]
  \centering
  \begin{subfigure}{.9\textwidth}
\includegraphics[width=\linewidth]{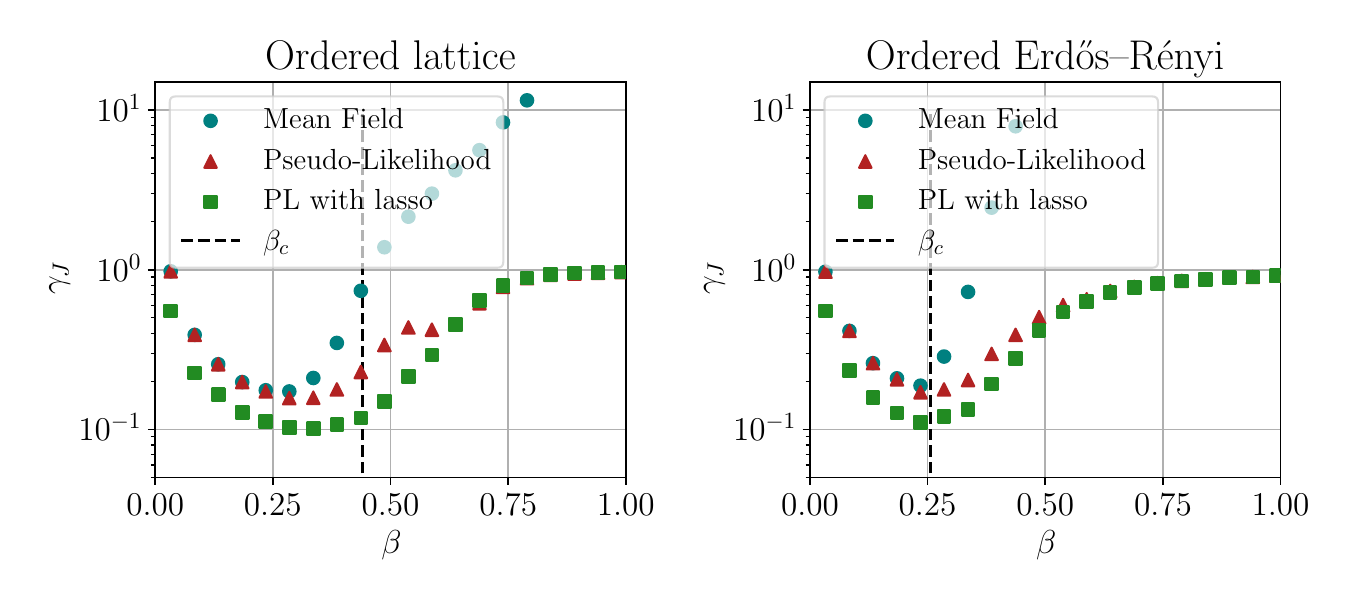}
  \end{subfigure}
  \hfill
  \begin{subfigure}{.9\textwidth}
\includegraphics[width=\linewidth]{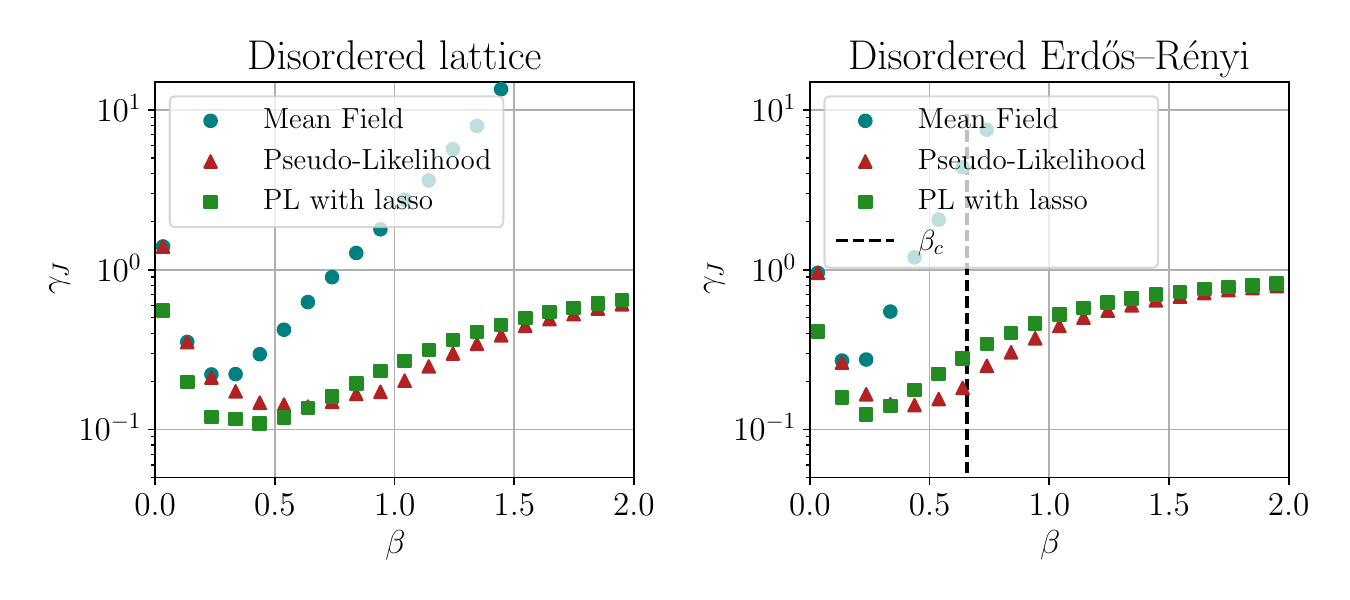}
  \end{subfigure}
  \caption{Reconstruction error $\gamma_J$ as a function of temperature for the Ising model ($8 \times 8$ square lattice and Erdős–Rényi graph with $N = 64$) using Mean-field, Pseudo-likelihood and LASSO regularization. Teal circles, red triangles and green squares are the errors obtained via the Mean Field, Pseudo-Likelihood and Pseudo-Likelihood with lasso (lasso parameter set to 1) methods, respectively. Black dashed vertical lines identify the inverse critical temperatures $\beta_c$ for the models. Data have been obtained using 20000 configurations obtained by independent Monte Carlo simulations. For the Pseudo-Likelihood results, a gradient descent with exponential decay of the learning rate was performed (parameters: starting learning rate 10, decay factor 0.999, epochs 5000).}
  \label{fig:griglia}
\end{figure}

As an example, the reconstruction error as a function of $T$ for four different kinds of  models with Ising spin variables is shown in Fig. \ref{fig:griglia} in the case of $N = 64$ spins. 
The models differ by the adjacencies of the spins and by the values of the couplings exchanged between adjacent spins.

In the top panels we present the ferromagnetic model on a square lattice and on an ER random graph of average connectivity $4$. The ferromagnetic models have all couplings equal to each other and positive. 
In the bottom panel of Fig. \ref{fig:griglia} the reconstruction errors for spin-glass models are presented, in which the values of the couplings are random variables distributed according to a Gaussian distribution of zero mean and unitary variance.
The networks are a square 2D lattice (with helicoidal boundary conditions) and an ER $G_N(p)$ graph of average connectivity $4$.

We notice that the Max-Pseudo-Likelihood approach of Sec. \ref{Sec:MAX_PL} performs equally or  better than the Mean Field one at every temperature.

$\mathrm{(}\star\mathrm{)}$ For those who looked at Sec. \ref{OVERFITTING}, about regularization, we also report the reconstruction error introducing the lasso regularization in the log Pseudo-Likelihood. Lasso further decreases the error obtained by means of the max PL in the ferromagnetic cases, in which there is a sharp difference between zero and non-zero couplings. In the non-ferromagnetic cases, on the other hand, introducing lasso does not noticeably improve the curves, yielding better results for some temperatures and worse results for others.


\subsubsection{$\mathrm{(}\star\mathrm{)}$ The Potts $4-$state clock model }

\begin{figure}[t!] 
    \centering
    \makebox[\textwidth][c]{\includegraphics[width=.9\textwidth]{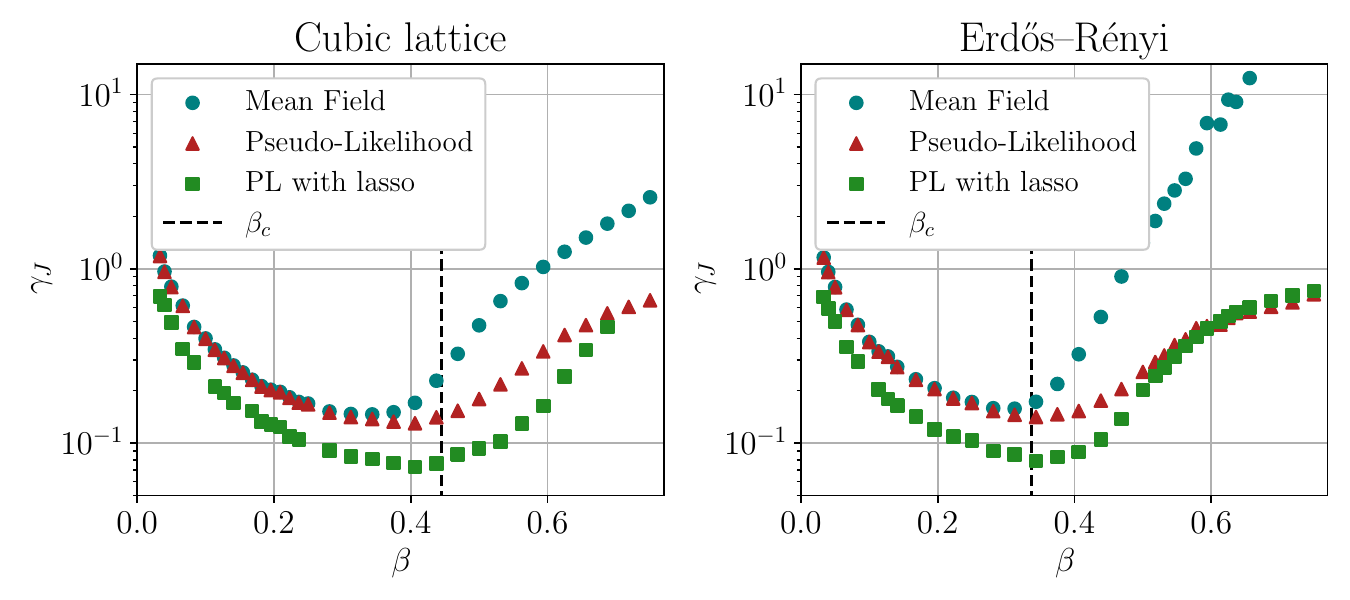}}%
    \caption{$\mathrm{(}\star\mathrm{)}$ Reconstruction error $\gamma_J$ as a function of temperature for the Potts model. \textit{Left}: 4x4x4 cubic lattice with periodic boundary conditions. \textit{Right}: Erdős–Rényi graph with $N = 64$ spins and average connectivity 6 (the same graph has been used for all data points). Teal circles, red triangles and green squares are the errors obtained via the Mean Field, Pseudo-Likelihood and Pseudo-Likelihood with lasso (lasso parameter set to 0.01) methods, respectively. Dashed vertical lines identify the inverse critical temperatures $\beta_c$ for the two models. Data have been obtained using 15000 Wolff steps by using one every 100 steps after thermalization. For the Pseudo-Likelihood results, a gradient descent with exponential decay of the learning rate was performed (parameters: starting learning rate 1, decay factor 0.999, epochs 5000).}
    \label{fig:both_potts}
\end{figure}

Similarly, the reconstruction error as a function of $T$ for the Potts model is shown in Fig. \ref{fig:both_potts} in the case of $N = 64$ spins for the vectorial Potts model with $q=4$. 
We considered a cubic 3D lattice (with helicoidal boundary conditions) and a $G_N(M)$ ER graph (this time with the number of links fixed) of average connectivity 6. In both cases we considered ferromagnetic interactions with $J = 1$.

Again, notice that the  Max-Pseudo-Likelihood approach of Sec. \ref{Sec:MAX_PL} performs equally or  better than the naive Mean Field one (see Sec. \ref{Sec:NMF}) at every temperature and lasso (Sec. \ref{OVERFITTING} further improves on these results, since we only consider ferromagnetic couplings in this case.

\subsubsection{$\mathrm{(}\star\mathrm{)}$ The Blume-Capel model }

\begin{figure}[t!] 
    \centering
    \makebox[\textwidth][c]{\includegraphics[width=1\textwidth]{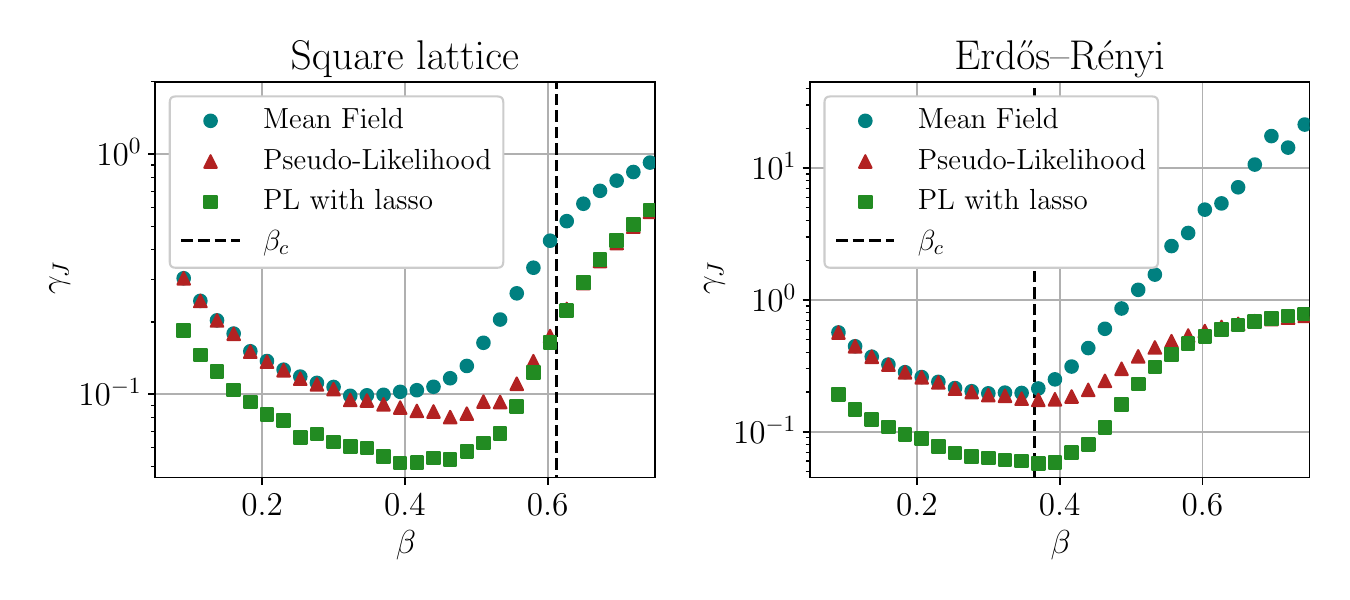}}%
    \caption{$\mathrm{(}\star\mathrm{)}$ Reconstruction error $\gamma_J$ as a function of temperature for the Blume-Capel model. \textit{Left}: $L=8$ square lattice with helicoidal boundary conditions. \textit{Right}: Erdős–Rényi graph with $N = 64$ spins and average connectivity 4 (the same graph has been used for all data points). Teal circles, red triangles and green squares are the errors obtained via the Mean Field, Pseudo-Likelihood and Pseudo-Likelihood with lasso (lasso parameter set to $10^{-4}$) methods, respectively. Dashed vertical lines identify the inverse critical temperatures $\beta_c$ for the two topologies. Each point in the graph is given by a dataset of 20000 independent samples obtained from a Parallel Tempering simulation. For the Pseudo-Likelihood results, a gradient descent with exponential decay of the learning rate was performed. In this case, the learning rate decay was applied once every 20 epochs (parameters: starting LR 10, decay factor 0.9, epochs 300).
    }
    \label{fig:bc_gammas}
\end{figure}

Finally, the results for the reconstruction error behavior with the data acquired at different temperatures  for the Blume-Capel model are shown in Fig. \ref{fig:bc_gammas}.
For this model we considered a square 2D lattice (with helicoidal boundary conditions) and a $G_N(p)$ ER graph of average connectivity 4. In both cases we considered ferromagnetic interactions with $J = 1$.

\begin{figure}[t!] 
    \centering
    \makebox[\textwidth][c]{\includegraphics[width=1\textwidth]{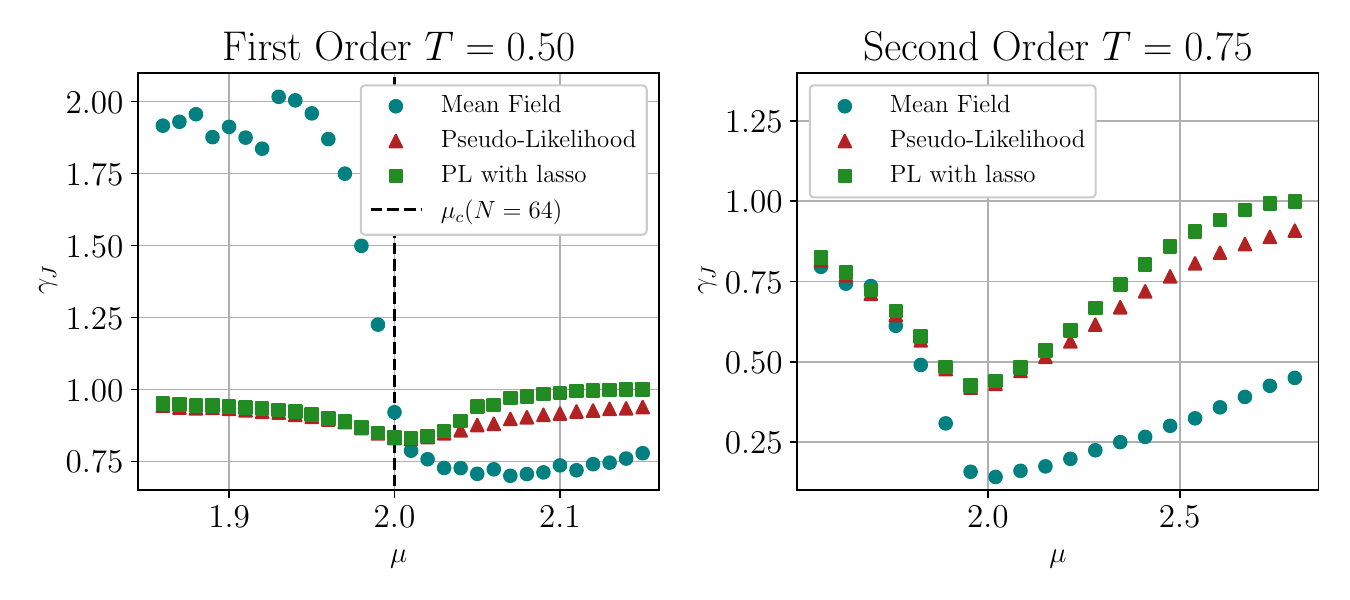}}%
    \caption{$\mathrm{(}\star\mathrm{)}$ Reconstruction error $\gamma_J$ as a function of temperature for the Blume-Capel model for a 2D lattice with $L=8$. \textit{Left}: Fixed $T = 0.5$ we vary $\mu$ to face a first order transition ar $\mu \approx 2$. \textit{Right}: Fixed $T = 0.75$ we vary $\mu$ to cross the second order transition line. Teal circles, red triangles and green squares are the errors obtained via the Mean Field, Pseudo-Likelihood and Pseudo-Likelihood with lasso (lasso parameter set to $10^{-4}$) methods, respectively. Dashed vertical lines identify the critical chemical potential $\mu_c$ for the left-side plot. Each point in the graph is given by a dataset of 20000 independent samples obtained from a Parallel Tempering simulation. For the Pseudo-Likelihood results, a gradient descent with exponential decay of the learning rate was performed. In this case, the learning rate decay was applied once every 20 epochs (parameters: starting LR 10, decay factor 0.9, epochs 300).}
    \label{fig:first_ord}
\end{figure}

In this particular case, where besides the second order phase transition also a first order phase transition occurs in a given region of the phase diagram, 
it is also interesting to see what happens to the reconstruction procedure when data are acquired in the region of parameters space where the first order transition takes place. 
In Fig. \ref{fig:bc_phasediag} we show that the Blume-Capel model undergoes a first-order transition for low temperature when the chemical potential $\mu\simeq 2$. 
The paramagnetic phase, occurring for  values of $\mu$ higher than the transition point, is originated by the dominant presence of $s_i = 0$ spins, that is of {\it holes} whose interaction contribution to the Hamiltonian is null.  This scenario is   fundamentally different from the one that happens at the second-order ferromagnetic transition, which is, instead, determined by thermally induced fluctuations of the $s_i= \pm 1$ spins and their  disalignment. Because of this, the inference in this region is hard since null spins give no information about the correlation. 
Indeed, if we look at the left plot of Fig. \ref{fig:first_ord}, we see that the mean-field reconstruction error is already high in the paramagnetic region $\mu>2$ but it increases sharply as it crosses the transition line at $T=0.5$. 
The Max-Pseudo-Likelihood error, on the other hand,  stays below $1$ but is not very far from it at any $\mu$, signaling that the procedure is unable to make a good reconstruction. For comparison, on the right panel of Fig. \ref{fig:first_ord} we show what happens when looking at the inference from data taken, once again,  at different $\mu$, but at a temperature $T=0.75$ at which the phase transition is second order.

\newpage

\subsection{Rank-plot analysis of the inferred couplings}
\label{RANK}
Another way of testing the  quality of the inference procedure is to observe the sorting of the values of the inferred couplings $J$ per decreasing value and also  to compare it with the same ordering for the original couplings.

\begin{figure}[t!]
    \centering
    \begin{subfigure}[b]{0.48\textwidth}
        \centering
        \includegraphics[width=\textwidth]{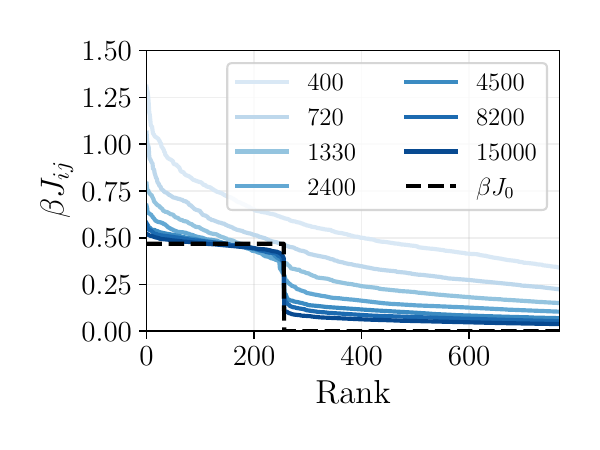}
        \caption{2D ordered Ising no lasso.}
    \end{subfigure}
    \hfill
    \begin{subfigure}[b]{0.48\textwidth}
        \centering
        \includegraphics[width=\textwidth]{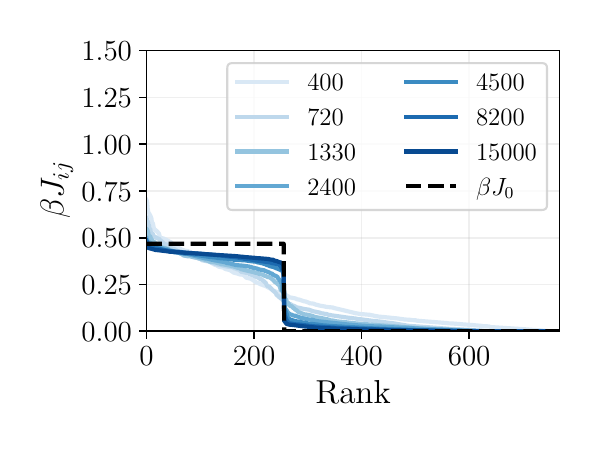}
        \caption{2D ordered Ising with lasso.}
    \end{subfigure}
    \caption{Values of the inferred coupling $\beta J_{ij}$ for the Ising ordered model on a square lattice, sorted from largest to smallest, as a function of the rank, the dashed line is the sorting original couplings $\beta J_{ij}$. $T = 2.14, \beta = 0.468$. (a): no lasso regularization has been used; (b): lasso regularization has been used. Data for $L = 8$ generation and inference have been performed using the same parameters as in Fig. \ref{fig:griglia}.}\label{fig:Jrank1}
\end{figure}

\begin{figure}[t!]
    \centering
    \begin{subfigure}[b]{0.48\textwidth}
        \centering
        \includegraphics[width=\textwidth]{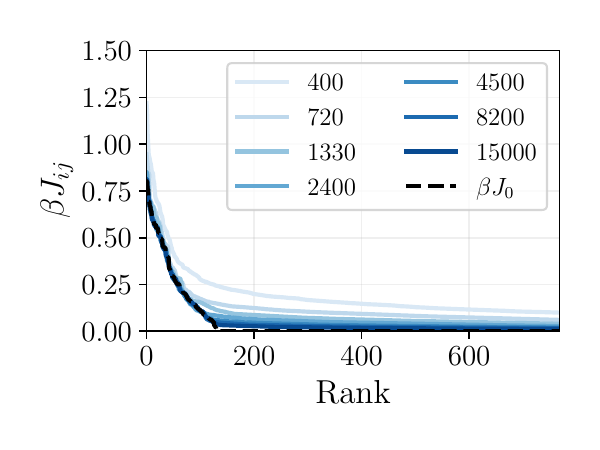}
        \caption{2D random-bond Ising no lasso.}
    \end{subfigure}
    \hfill
    \begin{subfigure}[b]{0.48\textwidth}
        \centering
        \includegraphics[width=\textwidth]{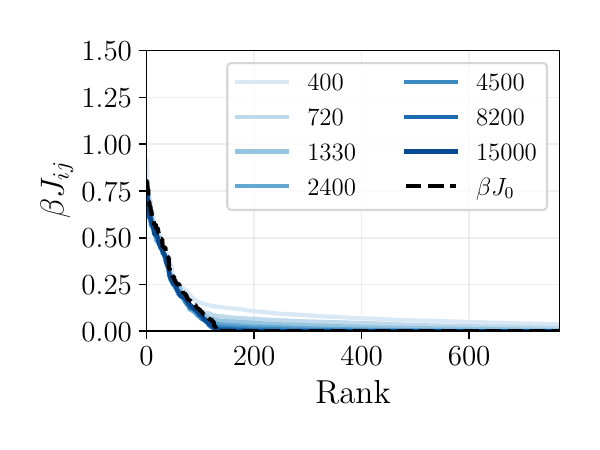}
        \caption{2D random-bond Ising with lasso.}
    \end{subfigure}
    \caption{Values of the inferred coupling $\beta J_{ij}$ for the Ising disordered model on a square lattice, sorted from largest to smallest, as a function of the rank, the dashed line is the sorting original couplings $\beta J_{ij}$. Only the region in which couplings are greater than 0 is shown. The high-rank region, in which the couplings assume negative values, is not show, but has a similar (although) mirrored behaviour with respect to the low-rank region. $T = 2.14, \beta = 0.468$. (a): no lasso regularization has been used; (b): lasso regularization has been used. Data for $L = 8$ generation and inference have been performed using the same parameters as in Fig. \ref{fig:griglia}.}\label{fig:Jrank1.1}
\end{figure}

\begin{figure}[t!]
    \centering
    \begin{subfigure}[b]{0.48\textwidth}
        \centering
        \includegraphics[width=\textwidth]{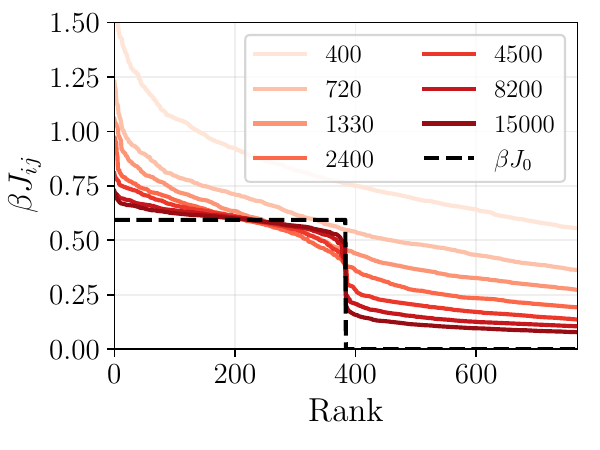}
        \caption{3D ordered $4$-clock model no lasso.}
    \end{subfigure}
    \hfill
    \begin{subfigure}[b]{0.48\textwidth}
        \centering
        \includegraphics[width=\textwidth]{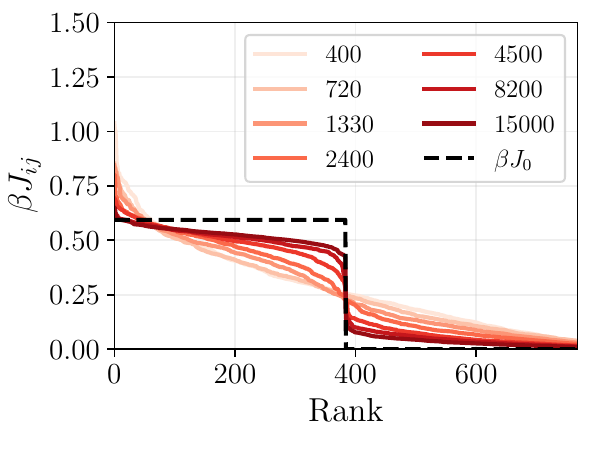}
        \caption{3D ordered $4$-clock model with lasso.}
    \end{subfigure}
    \caption{$\mathrm{(}\star\mathrm{)}$ Values of the inferred coupling $\beta J_{ij}$ for the Potts model on a 3D lattice, sorted from largest to smallest, as a function of the rank, the dashed line is the sorting original couplings $\beta J_{ij}$. $T = 1.68, \beta = 0.60$. (a): no lasso regularization has been used; (b): lasso regularization has been used. Data generation and inference have been performed using the same parameters as in Fig. \ref{fig:both_potts}.}\label{fig:Jrank2}
\end{figure}

\begin{figure}[t!]
    \centering
    \begin{subfigure}[b]{0.48\textwidth}
        \centering
        \includegraphics[width=\textwidth]{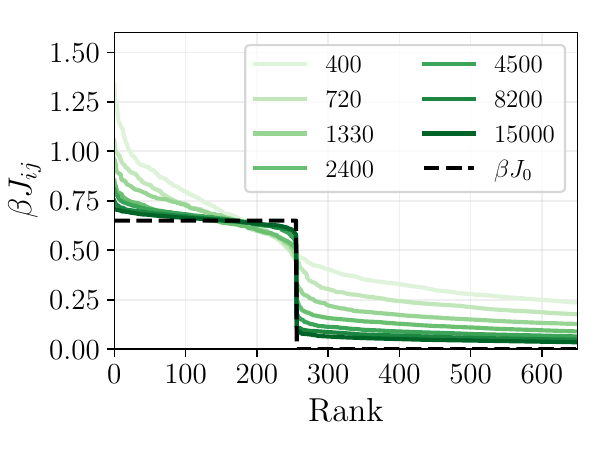}
        \caption{2D ordered Blume-Capel model no lasso.}
    \end{subfigure}
    \hfill
    \begin{subfigure}[b]{0.48\textwidth}
        \centering
        \includegraphics[width=\textwidth]{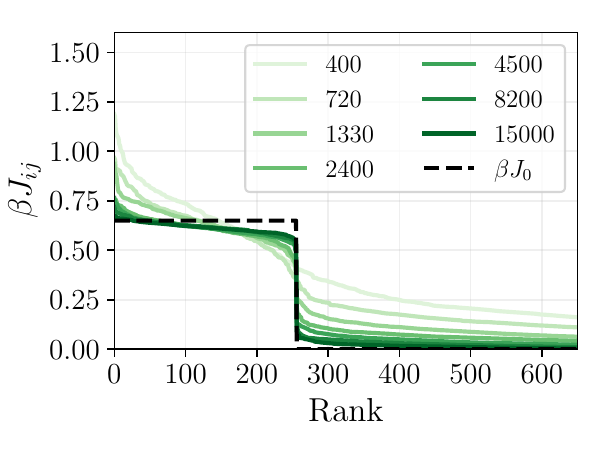}
        \caption{2D ordered Blume-Capel model with lasso.}
    \end{subfigure}
    \caption{$\mathrm{(}\star\mathrm{)}$ Values of the inferred coupling $\beta J_{ij}$ for the Blume Capel model on a 2D lattice, sorted from largest to smallest, as a function of the rank, the dashed line is the sorting original couplings $\beta J_{ij}$. $T = 2.40, \beta = 0.42$, $\mu = 0.25$. (a): no lasso regularization has been used; (b): lasso regularization has been used. Data generation and inference have been performed using the same parameters as in Fig. \ref{fig:bc_gammas}.}\label{fig:Jrank3}
\end{figure}

In Figs. \ref{fig:Jrank1} (the introductory track reader can focus on the Ising model case displayed in this figure), we plot the sorting of the couplings for different sizes of the experimental datasets acquired from  the same models as before, with and without a lasso regularizer. In the ferromagnetic models, a perfect inference would yield a step function behavior. In practice, the curves are  smoothed out, but the behavior  becomes sharper an sharper increasing the size of the dataset. 

$\mathrm{(}\star\mathrm{)}$ If you have been through Sec.  \ref{OVERFITTING}, the addition of a lasso regularization decreases the absolute values of the inferred couplings, and this translates in shifted-down curves for all the models. 
For the advanced readers $\mathrm{(}\star\mathrm{)}$, in figures \ref{fig:Jrank1.1},\ref{fig:Jrank2} and \ref{fig:Jrank3} we also present the same analysis carried out in vectorial Potts and Blume-Capel model instances.

\newpage

\newpage

\newpage 

\section{Conclusions}

In these paper we dedicated  several sections to give a concise, though hopefully clear enough pedagogic  overview of the statistical mechanics treatment of inverse problems in presence of phase transitions.
We divided the contents of the paper in an introductory track and in in-depth advanced sections denoted by  
$\mathrm{(}\star\mathrm{)}$.

In the first  part  (Sec. \ref{Sec:II}-\ref{INVERSE_MF}), we started by recalling the definition of the direct problem in the statistical mechanics sense, i.e., sampling equilibrium configurations of a physical system (e.g. with respect to the Gibbs-Boltzmann measure). We  focused in the Ising ordered and disordered models, but we also gave several details about  the vector Potts model and the Blume-Capel model in the advanced sections. We described how  the direct problem translates into the inverse one: reconstructing the probability distribution of the model starting from data. We, then,  detailed how this inverse problem can be tackled using the instruments of statistical mechanics, describing in particular the Maximum-Likelihood, Maximum-Pseudo-Likelihood and Mean-Field approaches. 

In the last part of the paper (Sec. \ref{Sec:SI}) we showed how these techniques can be applied in practice in  statistical mechanics systems undergoing phase transitions (both of the first or second order). We studied extensively the different systems earlier introduced and we considered these models both on regular lattices and on random graphs. We displayed the  performances of different inference procedures to reconstruct the interaction matrix starting from data. We highlighted how the testing of the inference procedure can be carried out either by considering the reconstruction error $\gamma_J$ as a function of the temperature of the data set, or by looking at the rank plots of the inferred couplings. In this way, we gave a general idea of how the reconstruction procedure behaves for interacting systems undergoing phase transition both close to and far away from criticality.

The presentation of the  numerical experiments is accompanied by a GitHub directory that allows direct reproducibility and implementation of the different procedures. This work serves not only as a theoretical introduction to inverse problems but also as a practical tool for hands-on learning, enabling readers to both reproduce results and apply the methods to new systems.

\section*{Code availability}
The code used in this paper is available at the GitHub repository \url{https://github.com/bsfn-0323/inverse_ising}.

\section*{Acknowledgements}
We thank Maria Chiara Angelini for useful discussions. We acknowledge  funding from the Italian Ministry of University and Research, call PRIN 2022, project “Complexity, disorder and fluctuations”, grant code 2022LMHTET.  This study was conducted using the HPC infrastructure DARIAH of the National Research Council of Italy, by the CNR-NANOTEC in Lecce,
funded by the MUR PON “Ricerca e Innovazione 2014-2020” program, project code PIR01-00022.

The authors have no conflicts to disclose.

\bibliography{biblio}
\newpage

\FloatBarrier
\appendix
\section{Estimate of the critical temperature}
\label{app:1}
True phase transitions only exist in the limit of an infinitely large system. However, simulations are run only for systems of finite size, even though state-of-the-art GPU codes simulate 2D Ising systems up to $L \lesssim 2^{23}$ \cite{bisson2025massive}. As a result, the samples obtained from simulations are affected by finite-size errors and exhibit a (more or less) different behavior from infinitely large systems. Since we are actually interested in computing observables in the thermodynamic limit, in order to derive the correct information on this quantities one performs the so-called finite size scaling analysis \cite{finitesizescaling}. Finite size scaling is a series of techniques aimed at extracting information on the infinite size limit of a system by taking measurements for systems of different (finite) sizes and then comparing them.\\
As an example, the critical temperature of a second order phase transitions is often obtained using the Binder parameter \cite{wenzel2011monte} \footnote{Other definitions are possible. For instance, in \cite{barone2006programmazione} it is defined as: 
\begin{equation}
    B(T) = \frac{1}{2}\Big(3- \frac{\langle m^4\rangle}{\langle m^2\rangle^2}\Big).
\end{equation}
}:
\begin{equation}
    B(T) = 1- \frac{\langle m^4 \rangle}{3\langle m^2\rangle^2} \\,
\end{equation}
where $m$ is the order parameter of the system and the average $\langle\cdot \rangle$ is over the Gibbs measure. 
The Binder parameter depends on both the temperature and the size of the system, but it becomes (for sufficiently large $N$) independent of the system size at the critical temperature (but only at that temperature!).
Therefore, by plotting $B(T)$ as a function of $T$ for different system sizes $L$ and finding the intersection point, it is then possible to estimate the critical temperature of the system. 

\subsection{Ising and Blume-Capel model}
\begin{figure}[t!]
\centering
\includegraphics[width = 1 \textwidth]{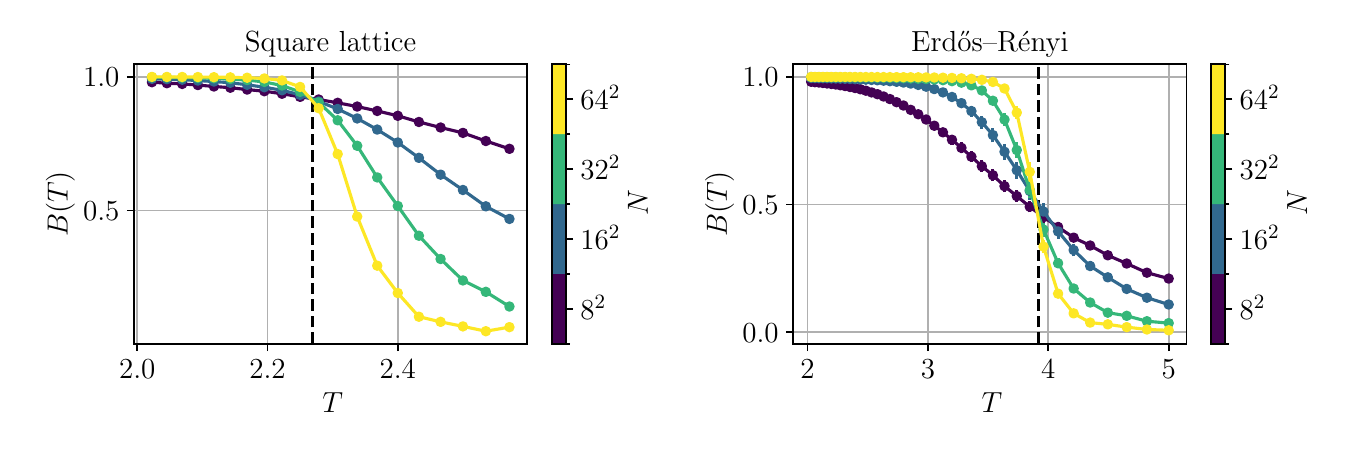}
\caption{\label{fig:binder_ising} Binder parameter as a function of temperature for different systems sizes (data obtained via Parallel Tempering) for the Ising model on a square lattice (\textit{left}) and on a ER graph $G_N(M)$ of average connectivity 4 (\textit{right}). Black dashed vertical lines correspond to the critical temperature of the models, $T_c \simeq 2.269$ for the square lattice and $T_c \simeq 3.9152$ for the ER graph. The value of the Binder parameter at each temperature has been obtained by averaging the results obtained for 10 different graphs. Lines are just a guide for the eye.}
\end{figure}

\begin{figure}[t!]
\centering
\includegraphics[width = 1 \textwidth]{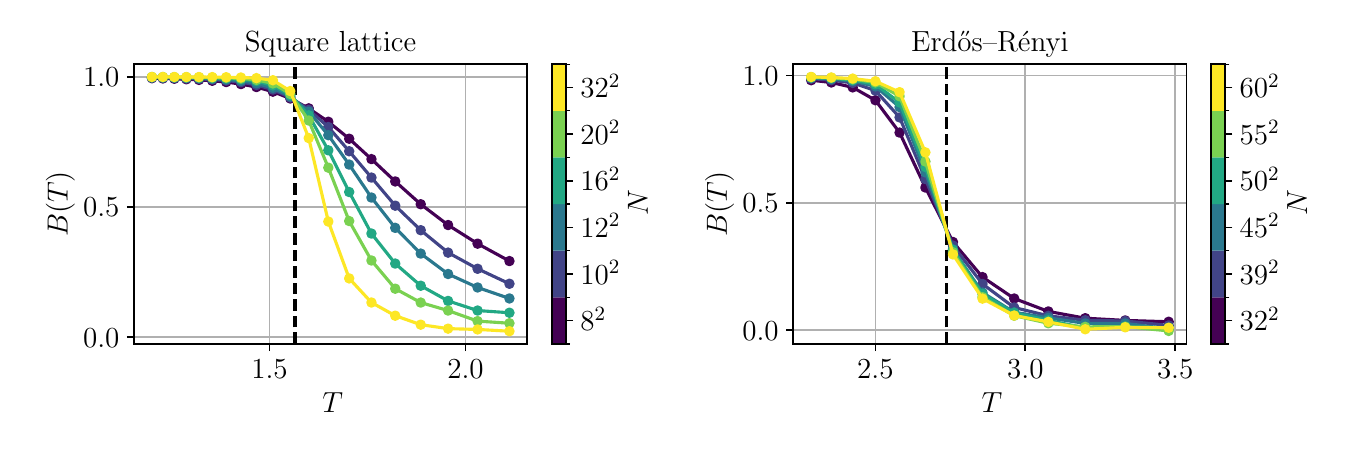}
\caption{\label{fig:binder_bc} Binder parameter as a function of temperature for different systems sizes (data obtained via Parallel Tempering) for the Blume-Capel model on a square lattice (\textit{left}) and on a ER graph $G_N(p)$ of average connectivity 4 (\textit{right}). Black dashed vertical lines correspond to the critical temperature of the models, $T_c(\mu=0.5) \simeq 1.565$ (cf. \cite{zierenberg2017scaling}) for the lattice and $T_c(\mu=0.25) \simeq 2.7376$ for the ER graph. The value of the Binder parameter at each temperature has been obtained by averaging the results obtained for 10 different graphs. Lines are just a guide for the eye. Error bars are smaller than data points.}
\end{figure}

The order parameter for the Ising and Blume-Capel is the magnetization:
\begin{equation}
    m = \frac{1}{N}\sum_{i=1}^{N} s_i.
\end{equation}
The only difference between the two models is that Ising variables can have value $s_i = \pm1$, while Blume-Capel spins can also have value zero. Since zero spin do not contribute to the magnetization, in the end they are described by the same order parameter, which will be $m\approx 0$ for high temperatures (due to the thermal noise) and $m\approx 1$ for lower temperatures.
The behavior of the binder parameter for the two models is shown in Fig. \ref{fig:binder_ising} and Fig. \ref{fig:binder_bc}: in both cases, the intersection of the curves identifies the critical temperature of the system.

\subsection{Potts model}\label{app:Tc_potts}
Due to the more complex nature of the variables in the Potts clock model, it is necessary to introduce a slightly different set of order parameters with respect to the Ising and Blume-Capel cases.

Indeed, for each color $c$, we define a magnetization $m_c$ as \cite{binder1981static}:
\begin{equation}
    m_c = f_c -\frac{\sum^q_{r \neq c} f_r}{q - 1} = \frac{f_cq-1}{q-1},
\end{equation}
where $f_c$ is the fraction of spins of color $c$.
In the $N \to \infty$ case, $m_c$ undergoes an abrupt change at the critical temperature $T_c$. In particular, even at finite system sizes, at $T<T_c$ all the spins tend to be aligned in the same direction (ferromagnetic phase), so that $f_c \approx 1$, $m_c \approx 1$ for that color and $f_c \approx 0$, $m_c \approx -\frac{1}{q-1}$ for all the others; for $T > T_c$, spins are aligned in random directions, therefore $f_c \approx \frac{1}{q}$ and $m_c \approx 0$ for all the colors (paramagnetic phase).

\begin{figure}[t!]
\centering
\includegraphics[width = 1 \textwidth]{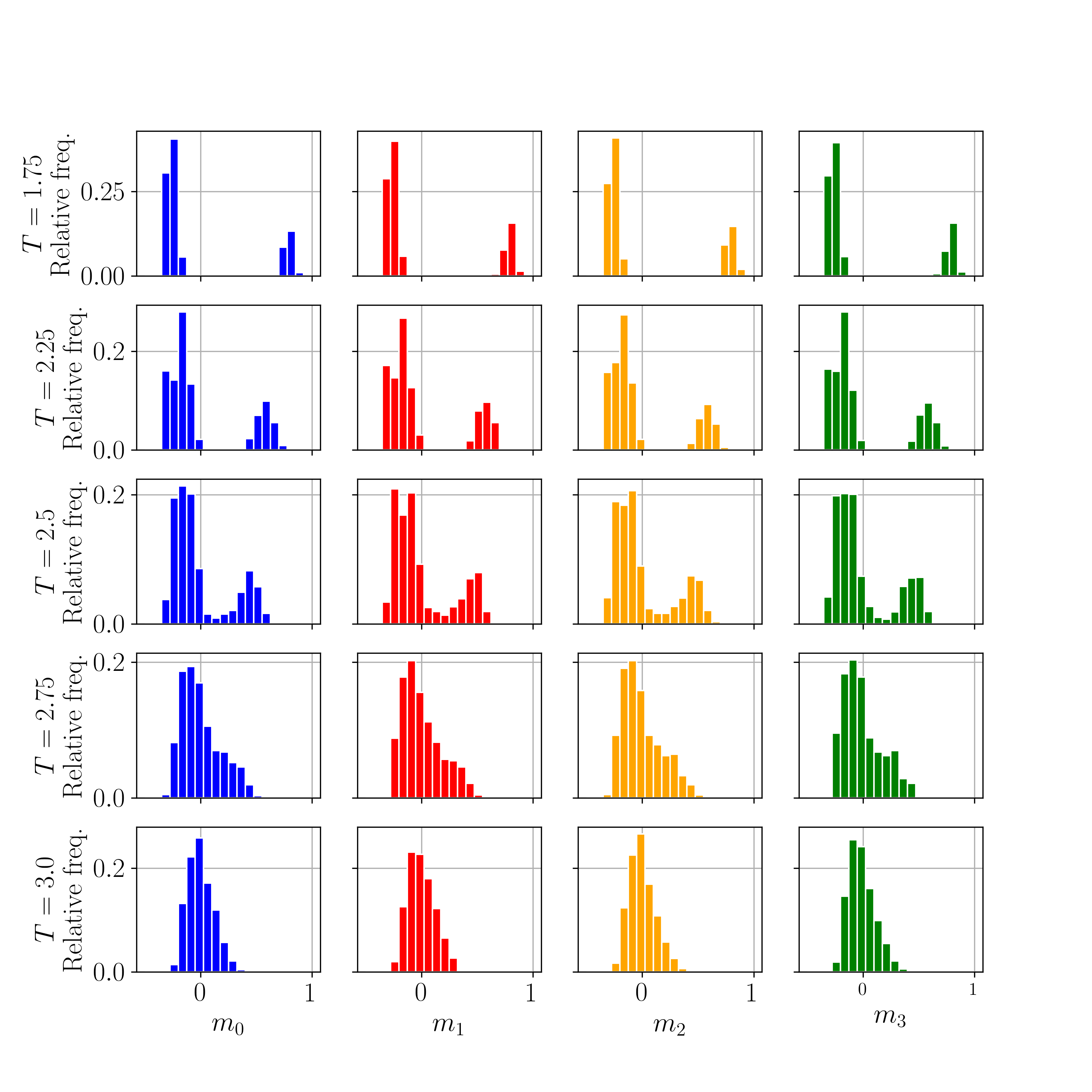}
\caption{ Histograms of the magnetizations of the four colors for different temperature obtained using the Wolff algorithm.}\label{fig:histo_wolff}
\end{figure}

In Fig. \ref{fig:histo_wolff} examples of magnetization histograms obtained for a system of $N= 216$ spins are shown. Data have been obtained using the Wolff algorithm \cite{wu2012critical}. 
Notice that the distribution is bimodal in the low-$T$, ferromagnetic phase, and it becomes unimodal in the high-$T$, paramagnetic phase.

\begin{figure}[t!]
    \centering
    \includegraphics[width=1\linewidth]{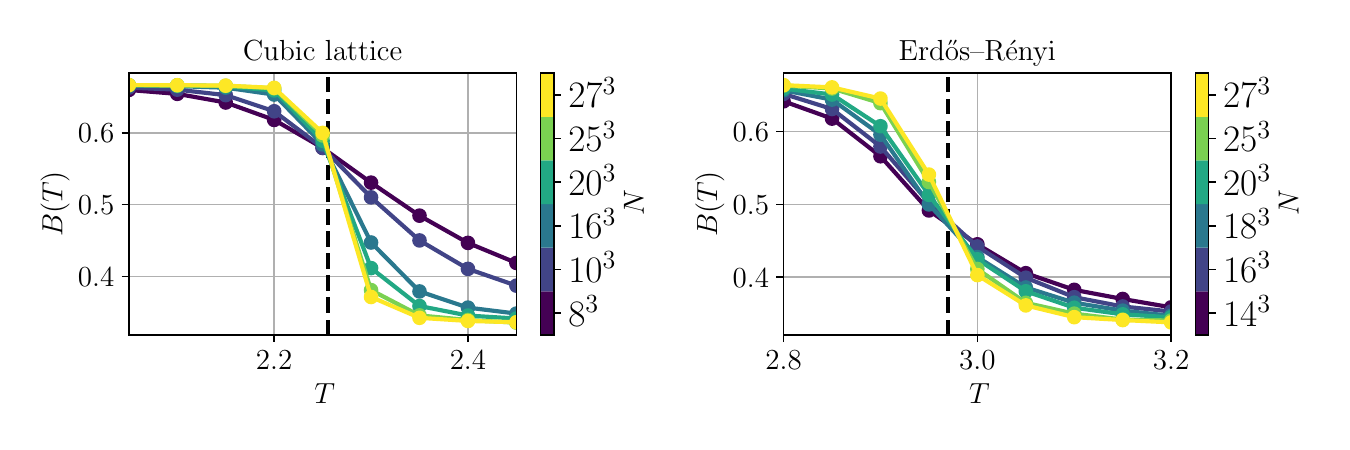}
    \caption{\label{fig:bc} Binder parameter as a function of temperature for different systems sizes (data obtained via the Wolff algorithm) for the Potts model on a cubic lattice (\textit{left}) and on a ER graph $G_N(p)$ of average connectivity 6 (\textit{right}). Black dashed vertical lines correspond to the critical temperature of the models, $T_c \simeq 2.26$ (cf. \cite{scholten1993critical}) for the cubic lattice and $T_c \simeq 2.97$ for the ER graph. The value of the Binder parameter at each temperature has been obtained by averaging the results obtained for 10 different graphs. Lines are just a guide for the eye. Error bars are smaller than data points.}
    \label{fig:binder_potts}
\end{figure}

To find the critical temperature we can introduce a parameter that characterizes the total configuration of the system \cite{wenzel2011monte, wu2012critical}:

\begin{equation}
    m = \frac{1}{N}\sqrt{(\sum_i \cos \theta_i)^2 + (\sum_i \sin \theta_i)^2},
\end{equation}
which, in the 4-colors case, can be rewritten as
\begin{equation}
    m = \sqrt{(f_0-f_2)^2 + (f_1-f_3)^2}.
\end{equation}
This procedure has been carried out in Fig. \ref{fig:binder_potts}. 

\newpage
\subsection{Spin–glass order parameter on an Erd\H{o}s–R\'enyi graph.}
\begin{figure}[t!]
\centering
\includegraphics[width = 0.6\textwidth]{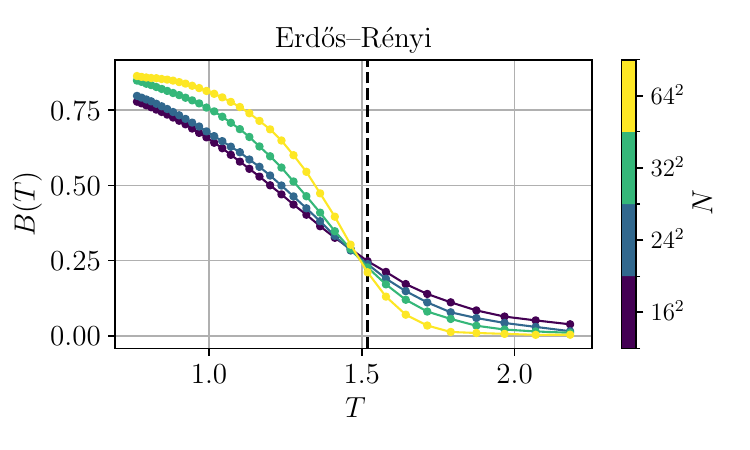}
\caption{\label{fig:binder_ising_random} Binder parameter as a function of temperature for different systems sizes (data obtained via Parallel Tempering) for the Ising model on a ER graph with Gaussian couplings. The black dashed vertical line correspond to the critical temperature of the models, $T_c = 1.524 $ \cite{viana1985phase}. The value of the Binder parameter at each temperature has been obtained by averaging the results obtained for 100 different realizations of graphs and couplings. The obtained critical temperature is slightly underestimated with respect to its theoretical value, probably due to finite-size effects, which are more pronounced in the disordered case and due to the fact that a $G_N(M)$ ensemble was used: at finite sizes, this differs from the $G_N(p)$ ensemble (e.g. in the free energy by a quantity of order $1/\sqrt{N}$ \cite{starr2008some}).  Lines are just a guide for the eye. Error bars are smaller than data points.}
\end{figure}

For the Ising model with zero‑mean Gaussian couplings on an Erd\H{o}s–R\'enyi random graph—the
\emph{Viana–Bray model}~\cite{viana1985phase}—geometric frustration forbids any ferromagnetic
ordering.  The conventional magnetization
\(
m=\tfrac1N\sum_{i=1}^N\langle s_i\rangle
\)
therefore vanishes identically at every temperature and cannot serve as an
order parameter.  
Instead, one monitors the \emph{Edwards–Anderson (replica) overlap} \cite{parisi1983order}
\begin{equation}
  q_{EA} \;=\;
  \Bigl\langle \frac1N\sum_{i=1}^{N}
  s_i^{(1)}\,s_i^{(2)} \Bigr\rangle_{J},
\end{equation}
where $(1)$ and $(2)$ denote two equilibrium configurations of the
\emph{same} disorder realization and $\langle\cdots\rangle_{J}$ includes the
average over couplings.
Above the critical temperature $T_c$ the overlap distribution $P_J(q)$ is sharply
peaked at $q=0$, whereas below $T_c$ it broadens and splits into multiple
peaks, signaling Replica Symmetry breaking and a genuine Spin‑Glass
phase \cite{mezard1987spin}.
In Fig. \ref{fig:binder_ising_random} is reported the Binder parameter for different system sizes. The intersections show a dependence on the system size, in particular, the obtained critical temperature seems to be slightly underestimated with respect to its theoretical value, probably due to finite-size effects, which are more pronounced in the disordered case. 
\newpage
\section{Numerical estimate of the critical crystal field}
\label{app:2}
In order to identify the first-order phase transition of the Blume-Capel model, one can monitor the probability distribution function $P_{T,\mu}(\rho)$ of the occupancy density $\rho$, i.e., the fraction of sites with nonzero spin:
\begin{equation}
    \rho = \frac{1}{N}\sum_{i=1}^N s_i^2.
\end{equation}

\begin{figure}[t!]
    \centering
    \includegraphics[width =0.7\textwidth]{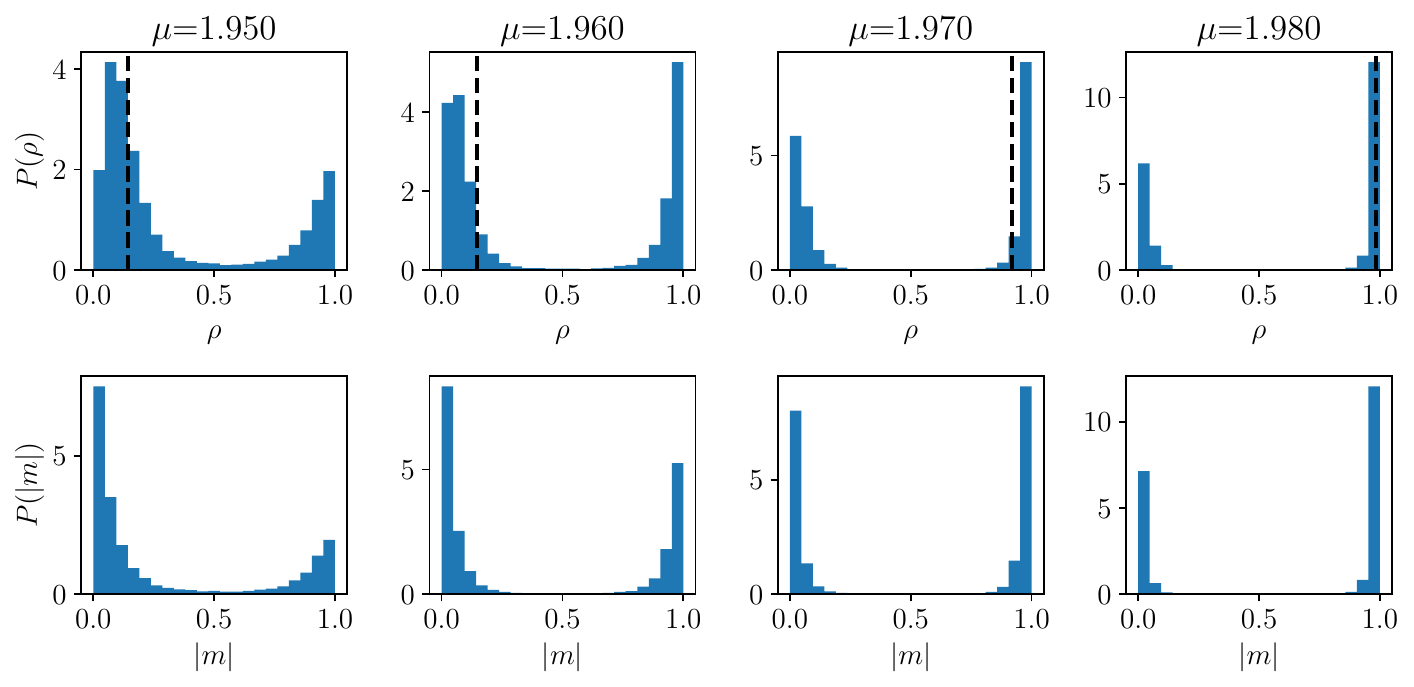}
\caption{$P(\rho)$ and $P(|m|)$ for a 2D lattice with $L=8$. First row: $P(\rho)$ at different values of $\mu$. The vertical dashed line is the median $\rho_0$. The critical value of $\mu$ will be $\mu_c \in (1.960,1970)$. This also can be seen in the second row where we plot the histogram of the absolute value of $m$. Between $\mu \in (1.960,1.970)$ the ferromagnetic peak becomes higher.}
    \label{fig:PTRho}

\end{figure}

Fig.~\ref{fig:PTRho} illustrates how $P_{T,\mu}(\rho)$ changes as $\mu$ varies (at a fixed temperature $T$). In particular, we observe two phases:
\begin{itemize}
  \item a \textbf{ferromagnetic Phase:} for smaller values of $\mu$, the distribution $P_{T,\mu}(\rho)$ peaks near $\rho \approx 1$, indicating that most spins occupy non-zero values and the system is magnetized;
  \item a \textbf{paramagnetic Phase:} as $\mu$ becomes sufficiently large, the peak shifts toward $\rho \approx 0$, meaning the system is predominantly in zero-spin states.
\end{itemize}

The transition between these two phases is a \textit{first-order transition}, at variance with the one that we described in the previous section, which in turn are \textit{second-order transitions}. At this first-order transition, the system displays \emph{phase coexistence}, reflected by a \emph{bimodal} $P_{T,\mu}(\rho)$: one peak near $\rho \approx 1$ and one near $\rho \approx 0$. To locate the critical chemical potential $\mu_c$ for that temperature $T$, one finds the value of $\mu$ for which the areas under the two peaks are equal. Equivalently, one can define $\rho_0$ (the ``median'' of the distribution) by the relation:
\begin{equation}
\int_{0}^{\rho_0} P_{T,\mu}(\rho) \, d\rho 
\;=\;
\int_{\rho_0}^{1} P_{T,\mu}(\rho) \, d\rho.
\end{equation}
When $\rho_0 = 1/2$,  the paramagnetic ($\rho \approx 0$) and ferromagnetic ($\rho \approx 1$) phases occur with the same probability, so $(T,\mu_c)$ is the point of coexistence for the first-order transition.

\section{Belief Propagation}\label{sec:BP}
\begin{figure}
    \centering
    \includegraphics[width=0.5\linewidth]{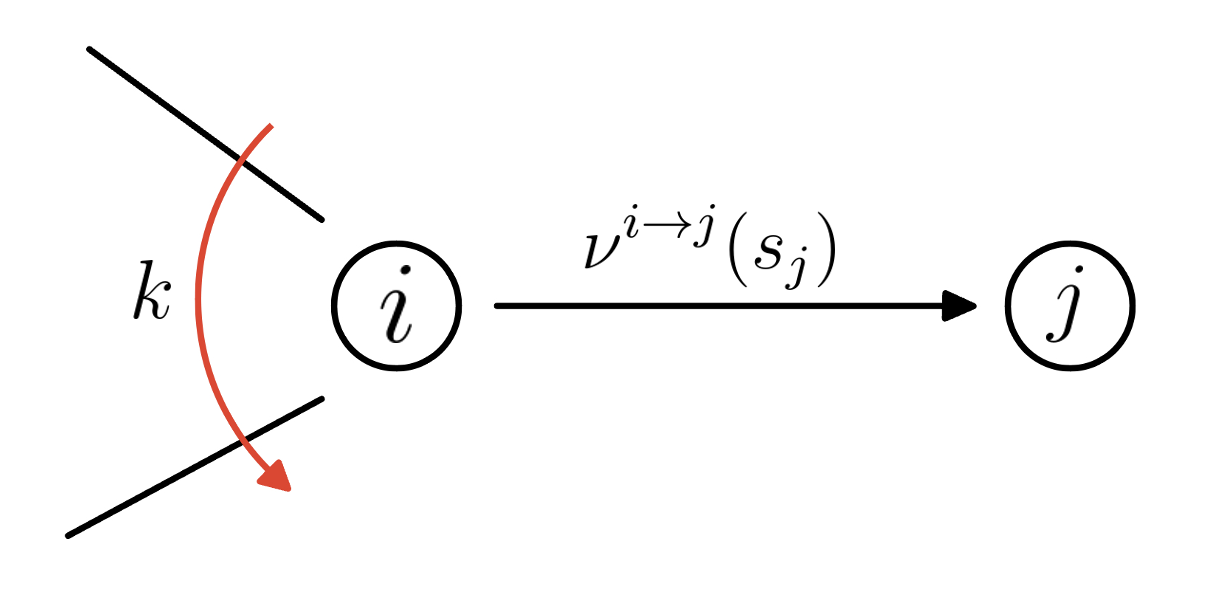}
    \caption{Sketch of the message entering in the node $j$ from node $i$. The neighbors of $i$, excluded $j$, are indexed by $k$.}
    \label{fig:sketch}
\end{figure}
To obtain the critical line of models defined on an Erdos-Renyi graph, one can use the Belief Propagation (BP) equations. Since the main purpose of this work is not to discuss the technicalities of this method, the interested reader can refer to Ref.  \onlinecite{mezard2009}. \\
In general with BP equations one describes the "message" from a node $i$ to a neighboring node $j$, $\nu^{i\rightarrow j}(s_i)$ (see Fig. \ref{fig:sketch}). One can consider it as the information that the node $j$ receives from node $i$. The main idea is to write simple self-consistent equations for the messages.\\ These equations can be written as
\begin{equation}
    \nu^{i \to j}(s_j) = \frac{1}{Z^{i \to j}} \psi_j^F (s_j) \sum_{\{s_i\}} \psi_{ij}^I (s_i, s_j) \prod_{k \in \partial i \setminus j} \nu^{k \to i} (s_i)
\end{equation}
where $\psi_i^F (\sigma_i)$ and $\psi_{ij}^I (\sigma_i, \sigma_j)$ are the field on node $i$ and the interaction term between $(i,j)$, while $Z^{i\rightarrow j}$ is the normalization. The product runs on all the neighbors of $i$ except $j$ as in Fig. \ref{fig:sketch}.\\
In models where all spins interact in the same way with other spins we can take all the messages to be equal, obtaining the simplified (single) equation for a connectivity $c$ and residual degree $d = c-1$:
\begin{equation}
    \nu(s) = \frac{1}{Z_{\nu}} \psi^F(s)  \sum_{\{s'\}} \psi^I (s', s) [\nu(s')]^d
    \label{eqn:mess}
\end{equation}
The above equations is correct for models on Random Regular (RR) graphs, in which each spin interacts with a fixed number of randomly chosen other spins. Strictly, they are not correct for ER graphs. However, when a second order transition happens both RR and ER have the same critical temperature, so we can use the above equation to obtain it nonetheless.
The corresponding single-spin marginal probability is then given by:
\begin{equation}
    p(s)= \frac{1}{Z_p} \psi^F(s)[\nu(s)]^d
    \label{eqn:marg}
\end{equation}
Equation \ref{eqn:mess} is easily solved: one simply iterates it until it converges to a final $\nu^*(s)$ and then use this value to compute the corresponding marginal via \ref{eqn:marg}. The magnetization will be given by \begin{equation}
    \langle s \rangle = \sum_{\{s\}}p(s)s
\end{equation}
The critical line (for the continuous transition described in this work) will be given by the values of the temperature for which the magnetization becomes non-zero.

For completeness, we give the explicit formulations of the self consistent equations for the models described in this work.
\subsection{Ising model}
For the Ising model, using the equation \ref{eqn:mess} one can find the following equations for the messages $\nu(\pm1)$:
\begin{align}
\nu(1) &= \frac{1}{\mathcal{Z}} \left[ e^{\beta}\nu(1)^d+e^{-\beta}\nu(-1)^d\right] \\
\nu(-1) &= \frac{1}{\mathcal{Z}} \left[e^{-\beta}\nu(1)^d+e^{\beta}\nu(-1)^d\right] \\
\label{eqn:ising_mess}
\end{align}
The reader can easily verify that these equations develop a non-trivial fixed point where $\nu(+1) \neq \nu(-1)$ at $T = T_c = 3.912$, in agreement with \eqref{Tc_ER_FM}.
\subsection{Blume-Capel Model}
For the Blume-Capel model, using the equation \ref{eqn:mess} one can find the following equations for the messages $\nu(0)$, $\nu(1)$ and $\nu(-1)$:
\begin{align}
\nu(0) &= \frac{1}{\mathcal{Z}} \left[\nu(0)^d+\nu(1)^d+\nu(-1)^d\right] \\
\nu(1) &= \frac{e^{-\beta \mu}}{\mathcal{Z}} \left[\nu(0)^d+ e^{\beta}\nu(1)^d+e^{-\beta}\nu(-1)^d\right] \\
\nu(-1) &= \frac{e^{-\beta \mu}}{\mathcal{Z}} \left[\nu(0)^d+e^{-\beta}\nu(1)^d+e^{\beta}\nu(-1)^d\right] \\
\label{eqn:bc_mess}
\end{align}
One re-obtains the solutions for the Ising Model by sending $\mu\rightarrow -\infty$\footnote{In practical terms, one has to iterate \ref{eqn:bc_mess} with a sufficiently large negative $\mu$.}. In this way the $s=0$ is suppressed. The reader can  verify that using these equations one recovers the critical line shown in Fig. \ref{fig:bc_phasediag}. Since these equation are true only for the RRG case, the first order transition line in Fig. \ref{fig:bc_phasediag} is only an approximation of the true one. 
\subsection{Potts Model}
For the Potts mode, calling $\nu(n)$ the message for color $n$, we find that the self consistent equations can be written as:
\begin{align}
\nu(0) &= \frac{1}{\mathcal{Z}} \left( e^{\beta J} \nu(0)^d + \nu(1)^d + e^{-\beta J} \nu(2)^d + \nu(3)^d \right), \\
\nu(1) &= \frac{1}{\mathcal{Z}} \left( \nu(0)^d + e^{\beta J} \nu(1)^d + \nu(2)^d + e^{-\beta J} \nu(3)^d \right), \\
\nu(2) &= \frac{1}{\mathcal{Z}} \left( e^{-\beta J} \nu(0)^d + \nu(1)^d + e^{\beta J} \nu(2)^d + \nu(3)^d \right), \\
\nu(3) &= \frac{1}{\mathcal{Z}} \left( \nu(0)^d + e^{-\beta J} \nu(1)^d + \nu(2)^d + e^{\beta J} \nu(3)^d \right).
\end{align}

where $\mathcal{Z}$ is a normalization required to ensure $\nu(1) + \nu(2) + \nu(3) + \nu(4) = 1$. The reader can easily verify that these equations develop a non-trivial fixed point at $T = T_c \simeq 2.97$, in agreement with Fig. \ref{fig:binder_potts}.


\end{document}